\documentclass[sigconf, authorversion=true]{acmart}
\AtBeginDocument{%
  \providecommand\BibTeX{{%
    \normalfont B\kern-0.5em{\scshape i\kern-0.25em b}\kern-0.8em\TeX}}}

\copyrightyear{2025}
\acmYear{2025}
\setcopyright{cc}
\setcctype{by}
\acmConference[CHI '25]{CHI Conference on Human Factors in Computing Systems}{April 26-May 1, 2025}{Yokohama, Japan}
\acmBooktitle{CHI Conference on Human Factors in Computing Systems (CHI '25), April 26-May 1, 2025, Yokohama, Japan}\acmDOI{10.1145/3706598.3714206}
\acmISBN{979-8-4007-1394-1/25/04}

\usepackage{graphicx}
\usepackage{stfloats}
\usepackage{float,acmart-taps}

\begin{document}

\title[Augmenting the Letter-Exchange Exercise with LLM-based Agents]{Letters from Future Self: Augmenting the Letter-Exchange Exercise with LLM-based Future Self Agents to Enhance Young Adults’ Career Exploration}

\author{Hayeon Jeon}
\orcid{0009-0003-5864-1185}
\affiliation{
    \institution{hci+d lab.}
  \institution{Seoul National University}
  \country{South Korea}
}
\email{jhy94520@snu.ac.kr}

\author{Suhwoo Yoon}
\orcid{0009-0004-5893-5439}
\affiliation{
    \institution{hci+d lab.}
  \institution{Seoul National University}
  \country{South Korea}
}
\email{yeopil@snu.ac.kr}

\author{Keyeun Lee}
\orcid{0009-0001-2794-5712}
\affiliation{
    \institution{hci+d lab.}
  \institution{Seoul National University}
  \country{South Korea}
}
\email{kieunp@snu.ac.kr}

\author{Seo Hyeong Kim}
\orcid{0009-0003-9069-7998}
\affiliation{
    \institution{hci+d lab.}
  \institution{Seoul National University}
  \country{South Korea}
}
\email{mariaksh@snu.ac.kr}

\author{Esther Hehsun Kim}
\orcid{0000-0002-9576-4411}
\affiliation{
    \institution{hci+d lab.}
  \institution{Seoul National University}
  \country{South Korea}
}
\email{ehk@snu.ac.kr}

\author{Seonghye Cho}
\orcid{0009-0006-9878-9952}
\affiliation{
    \institution{hci+d lab.}
  \institution{Seoul National University}
  \country{South Korea}
}
\email{sky4172@snu.ac.kr}

\author{Yena Ko}
\orcid{0009-0001-5158-9236}
\affiliation{
    \institution{Department of Communication}
  \institution{Seoul National University}
  \country{South Korea}
}
\email{kfree08@snu.ac.kr}

\author{Soeun Yang}
\orcid{0000-0003-2294-5403}
\affiliation{
    \institution{Department of Communication}
  \institution{Seoul National University}
  \country{South Korea}
}
\email{soeun022@snu.ac.kr}

\author{Laura Dabbish}
\orcid{0000-0001-9489-4188}
\affiliation{
    \institution{Human-Computer Interaction
Institute}
  \institution{Carnegie Mellon University}
  \country{United States}
}
\email{dabbish@cmu.edu}

\author{John Zimmerman}
\orcid{0000-0001-5299-8157}
\affiliation{
    \institution{HCI Institute}
  \institution{Carnegie Mellon University}
  \country{United States}
}
\email{johnz@cs.cmu.edu}

\author{Eun-mee Kim}
\orcid{0000-0003-4032-4731}
\affiliation{
    \institution{Department of Communication}
  \institution{Seoul National University}
  \country{South Korea}
}
\email{eunmee@snu.ac.kr}

\author{Hajin Lim}
\orcid{0000-0002-4746-2144}
\authornote{Corresponding Author}
\affiliation{
    \institution{hci+d lab.}
  \institution{Seoul National University}
  \country{South Korea}
}
\email{hajin@snu.ac.kr}

\renewcommand{\shortauthors}{Hayeon Jeon et al.}

\begin{abstract}
Young adults often encounter challenges in career exploration. Self-guided interventions, such as the letter-exchange exercise, where participants envision and adopt the perspective of their future selves by exchanging letters with their envisioned future selves, can support career development. However, the broader adoption of such interventions may be limited without structured guidance. To address this, we integrated Large Language Model (LLM)-based agents that simulate participants’ future selves into the letter-exchange exercise and evaluated their effectiveness. A one-week experiment (N=36) compared three conditions: (1) participants manually writing replies to themselves from the perspective of their future selves (baseline), (2) future-self agents generating letters to participants, and (3) future-self agents engaging in chat conversations with participants. Results indicated that exchanging letters with future-self agents enhanced participants' engagement during the exercise, while overall benefits of the intervention on future orientation, career self-concept, and psychological support remained comparable across conditions. We discuss design implications for AI-augmented interventions for supporting young adults' career exploration.
\end{abstract}



\maketitle

\section{Introduction}

\begin{figure*}[!ht]
    \centering
    \includegraphics[width=440pt]{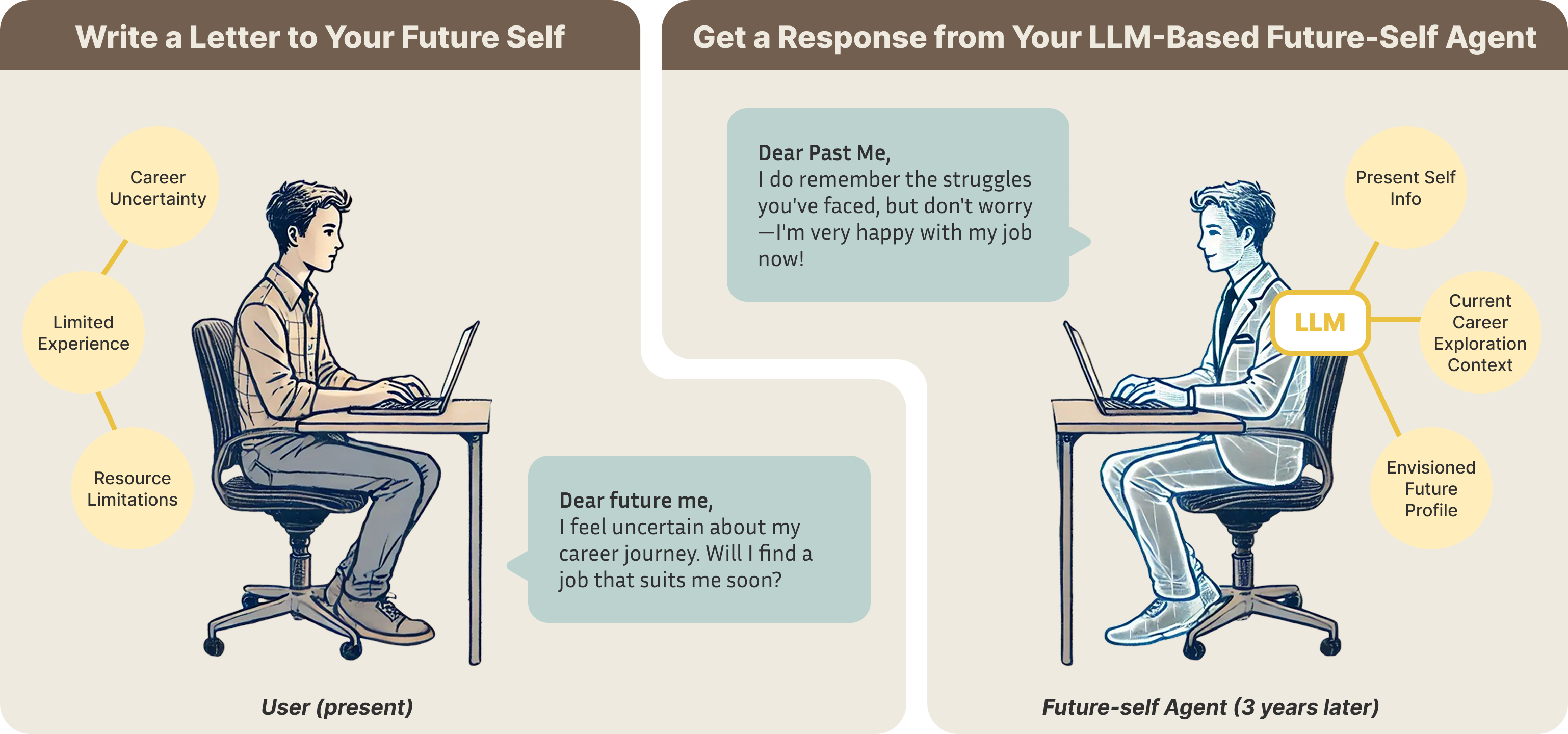}
    \caption{In this work, we investigate how LLM-based future-self agents can support young adults' career exploration. Participants write letters to their future selves (left). Then, their LLM-based future-self agents reply to their letters (right).}
    \Description{This is a teaser image for this entire paper. There are two boxes placed side by side. In the left box, the heading reads, "Write a Letter to Your Future Self." A student is typing on his laptop: "Dear future me, I feel uncertain about my career journey. Will I find a job that suits me soon?". This student is experiencing difficulties due to career uncertainty, limited experience, and resource limitations. In the right box, the heading reads, "Get a Response from Your LLM-Based Future-Self Agents." A future businessman version of the student is typing on his laptop: "Dear past me, I remember the struggles you faced, but don’t worry. I’m very happy with my job now!" This future version of the student is generated by an LLM, utilizing the student's present self info, current career exploration context, and envisioned future profile.}
    \label{fig:1}
\end{figure*}

Young adulthood is a pivotal stage characterized by significant life decisions, particularly in the realm of career choices \cite{arnett2015oxford}. Many individuals in this phase struggle to envision their future, hindering the development and clarification of their career paths \cite{oyserman2023possible, kwok2018managing}. This uncertainty often manifests as increased stress, anxiety, and diminished self-efficacy \cite{boo2020career, park2017mediation}. To address these challenges, researchers have developed various self-guided interventions to support young adults in their career pursuits \cite{soares2022systematic}, targeting aspects such as facilitating career information seeking \cite{teychenne2019pre,lam2018impact}, job search skill training \cite{chukwuedo2022practitioners}, and providing psychological support to alleviate stress and anxiety associated with career uncertainty \cite{ogbuanya2018effect}.

Recent studies have emphasized the importance of future orientation in facilitating young adults' career exploration \cite{chishima2021conversation}. One promising approach is \textbf{the letter-exchange exercise}, a self-guided writing intervention designed to promote future-oriented mindsets in the career pursuit process \cite{chishima2021temporal, chishima2021conversation}. This exercise comprises the `\textbf{Send Session},' where individuals write letters to their future selves, and the `\textbf{Reply Session},' where they respond to their present selves from the perspective of their future selves. By allowing young adults to create vivid futures and foster a connection with their future selves, this approach has proven effective in cultivating a future-oriented mindset, clarifying career goals, and enhancing psychological resilience \cite{kappes2014emergence, oettingen2012future, oettingen2000expectancy}. 

However, the cognitive demands of such self-directed activities can pose challenges for young adults, particularly when access to professional support is limited \cite{amanvermez2022effects, chishima2021temporal}.  Recent advancements in Large Language Model (LLM)-based conversational agents offer new possibilities for enhancing self-guided interventions by providing contextualized and personalized support \cite{shukuri_meta-control_2023, 10.1145/3544548.3581503}. Building on this, the present study aims to explore the potential of LLM agents that simulate participants' future selves, aligning with the core concept of communication between present and future selves in the letter-exchange exercise.

In the study, we conducted a one-week between-subjects experiment (N=36) to examine the impact of LLM agents simulating one's future self on young adults' career exploration. In the experiment, participants first completed the Send Session by composing letters to their future selves and proceeded to the Reply Session under one of three conditions: (1) \textbf{Writing Condition}, in which participants manually wrote replies to themselves from the perspective of their future selves as in the original letter-exchange exercise, (2) \textbf{LLM Letter Condition}, in which participants received letters generated by future-self agents, or (3) \textbf{LLM Chat Condition}, in which participants engaged in chat conversations with future-self agents. 

To assess the effectiveness of each condition, we employed a mixed-methods approach. This included collecting and analyzing quantitative measures before, immediately after, and one week following the exercise. Additionally, we collected and analyzed qualitative data from interviews to gain deeper insights into participants' perceptions of and experiences with the future-self agents. 

Building on these methods, our study aimed to address the following research questions (RQs):
\begin{itemize}
    \item \textbf{RQ1 (Participants’ Engagement):} How do interactions with LLM agents simulating one’s future self (LLM Letter and Chat Conditions) influence young adults’ \textbf{engagement during the letter-exchange exercise}?
    \item \textbf{RQ2 (Exercise Effectiveness):} How does the integration of LLM agents simulating one’s future self (LLM Letter and Chat Conditions) impact the \textbf{effectiveness of the letter-exchange exercise} in enhancing Connectedness with the Future Self, Career Goal Clarity, and Psychological Resilience?
    \item \textbf{RQ3 (Participants’ Experience):} How do \textbf{participants' experiences} differ across the three conditions in supporting young adults' career exploration process?
\end{itemize}

Our findings showed that integrating LLM agents into the letter-exchange exercise could effectively support young adults' career exploration, positioning LLM-based approaches as viable alternatives to the traditional self-guided methods. By transforming the solitary exercise into an interactive, guided experience with future-self agents, the LLM Letter Condition notably enhanced participants' engagement with the exercise itself  \textbf{(RQ1)}. Furthermore, the incorporation of LLM agents demonstrated effectiveness comparable to the conventional self-guided method \textbf{(RQ2)}, offering diverse pathways for career exploration \textbf{(RQ3)}. Specifically, while the Writing Condition promoted in-depth self-reflection, the LLM Letter Condition provided a contemplative and emotional experience through personalized letters, and the LLM Chat Condition facilitated dynamic conversations that supported active career exploration.

Drawing on these findings, we discuss design implications for AI-augmented career exploration, carefully weighing the opportunities and concerns of LLM technology in influencing young adults' self-reflection and decision-making processes as they navigate potential career paths.

Our contributions are as follows:
\begin{itemize}
    \item Developing a framework to implement LLM agents that simulate young adults' future selves to support their career exploration.
    \item Thoroughly examining the impact of LLM agents on young adults' career exploration from multiple angles, including engagement, effectiveness, and overall experience.
    \item Providing design implications for a balanced integration of LLM agents into young adults' career exploration that promotes active self-reflection while respecting user agency and privacy.
\end{itemize}


\section{Related Work}

Below, we first discuss the importance of young adulthood and career pursuits. Then, we introduce interventions designed to support these processes, focusing on the letter-exchange exercise and the associated accessibility and cognitive challenges in engaging with them without professional or structured support. Finally, we highlight the potential of LLM conversational agents in enhancing self-guided interventions for career exploration.

\subsection{Young Adulthood and Career Pursuits}

Young adulthood (ages 18 to 29) is a pivotal life stage that marks the beginning of adulthood, where an individual transitions into complete independence and confronts significant life choices \cite{arnett2000emerging}. During this period, young adults encounter a variety of life choices, ranging from educational decisions to financial security, romantic relationships, and community responsibilities \cite{mitra2021life}. Among these, career pursuits are regarded as one of the most critical decisions that young adults have to make \cite{arnett2023emerging}.

During the transition to adulthood, setting career objectives plays a crucial role in shaping young adults' identities and guiding the direction of their future lives \cite{masten2004resources,schulenberg2004taking,erikson1994identity}. The career decisions made at this developmental stage are known to have a lasting impact on their future lives \cite{hlavdo2022exploring,schulenberg2004taking,arnett2000emerging}. Also, career choices have a decisive influence on future well-being and overall life satisfaction \cite{macek2015emerging}.
To cultivate a rich and stable identity, it is essential for young adults to explore various career paths, then learn to navigate the challenges that arise during their career pursuits, and finally commit to their chosen profession \cite{marcia1966development}. 

Despite the importance of career pursuits, many young adults often struggle to clearly envision their future due to a lack of information, knowledge, and experience required to make informed career decisions \cite{gaffner2002factors}. Also, they often feel that their future still remains clouded in ambiguity and uncertainty \cite{kwok2018managing}. These vague visions of the future are known to significantly complicate young adults’ career development \cite{oyserman2023possible}. As a result, young adults who have not yet defined their career paths frequently experience substantial stress \cite{boo2020career}, leading to heightened anxiety and reduced self-efficacy \cite{park2017mediation,vignoli2015career,lazarus1984stress}.

To address this, a range of interventions have been developed to support young adults in their career pursuits \cite{soares2022systematic}. These include providing career-related information \cite{teychenne2019pre,lam2018impact}, guiding career exploration \cite{kepir2020effectiveness, pordelan2018online}, training job search skills \cite{chukwuedo2022practitioners}, and alleviating negative emotions \cite{ogbuanya2018effect}. Although these efforts cover important areas such as informational resources, skills development, and mental health support \cite{soares2022systematic}, they often overlook the importance of fostering \textit{future orientation}—a critical component in career guidance for young adults. Future orientation, also referred to as a future-oriented mindset, is an individual's ability to think about, connect with, and plan for their future \cite{chishima2021conversation}. In the context of career exploration, it entails envisioning long-term goals, identifying the steps required to achieve them, and aligning present actions with desired outcomes. This mindset is vital for young adults to navigate their career paths, as it enables them to transform existing knowledge and skills into meaningful progress toward their aspirations \cite{oyserman2023possible, kao2022see, guan2017modeling, strauss2012future}.

\subsection{The Letter-Exchange Exercise}
In this context, the \textit{letter-writing interventions}, a series of self-guided letter-writing exercises directed towards one's future self, have garnered significant interest as an effective means of cultivating future orientation \cite{barnett2019influence, rutchick2018future, van2013vividness}. While these interventions have shown promise in inducing future-oriented mindsets, the specific process varies across studies, including differences in time frames (e.g., 3 years vs. 20 years) and letter directionality (e.g., letter to the future self vs. from the future self) \cite{barnett2019influence}. 

Among such variations in letter-writing interventions, we specifically focused on the \textit{letter-exchange exercise}, a recently enhanced version of the traditional one-way letter-writing interventions, which expanded it to a bidirectional exchange of letters with one's future self \cite{chishima2021conversation}. This exercise was devised to further promote future orientation in young adults to support their career exploration. The exercise consists of two stages, each comprising two sub-activities: (i) the \textbf{Send Session}, where individuals first create a realistic profile of their future selves and then write letters to their future selves based on the created profile, and (ii) the \textbf{Reply Session}, where individuals first imagine their future selves and then write letters back to their present selves from this future perspective. To maximize the effectiveness of the exercise, it is recommended that the future self be positioned at a distance sufficient to represent the next step in career development yet close enough to be vividly imagined (e.g., 3 years) \cite{chishima2021conversation}.

The letter-exchange exercise proved its effectiveness in supporting young adults' career exploration by encouraging future orientation \cite{chishima2021conversation}. The bidirectional process facilitates young adults to simultaneously imagine their future selves from the present perspective while reflecting on their current selves from a future viewpoint. As a result, young adults gain clearer visions of their future and strengthen their connection to their future selves, thereby developing future-oriented mindsets. Subsequent studies have further validated its effectiveness by applying the exercise in a range of contexts \cite{pownall2024hope, warmath2024letter, chishima2021temporal}.

Despite its effectiveness in supporting young adults’ career pursuits, the letter-exchange exercise has limitations in facilitating participant engagement due to its self-guided nature \cite{10.1145/3613904.3642761, amanvermez2022effects}. Particularly, the exercise can be challenging for young adults to repetitively perform on their own over the long term, as it involves an intense cognitive process of comparing and contrasting two different temporal perspectives \cite{oettingen2012future, oettingen2000expectancy}. As a result, this process may be cognitively demanding for young adults to execute independently, particularly without professional guidance, which may hinder sustained engagement over time \cite{baumel2019objective, fleming2018beyond}. Additionally, individual differences in imaginative and writing capabilities may hinder the full realization of the intervention's effectiveness \cite{cheong2022role, chishima2021conversation, meevissen2011become}. While structured guidance from career professionals can be highly beneficial \cite{amanvermez2022effects, chishima2021temporal, chishima2021conversation}, limited access to such resources prevents many young adults from fully benefiting from the self-guided interventions, including the letter-exchange exercise.

\subsection{LLM-based Conversational Agents in Self-guided Interventions}

To address the limitations and barriers in self-guided interventions, conversational agents have emerged as powerful tools in delivering self-guided interventions across various domains, including mental health \cite{liu_using_2022, greer_use_2019, fitzpatrick_delivering_2017} and behavioral change \cite{gabrielli_chatbot-based_2020, bickmore2005establishing}. Early implementations of such agents were largely based on rule-based systems \cite{abdulrahman2023changing, ly_fully_2017, klein2014intelligent}, which exhibited significant limitations. Specifically, these systems offered limited adaptability to individual contexts and faced scalability challenges, as tailoring rules to each user was resource-intensive. Consequently, these limitations hindered their broader adoption and reduced their overall effectiveness \cite{martins_unlocking_2024}.

In this regard, the introduction of Large Language Models (LLMs) has the potential to enhance the capabilities of conversational agents, addressing the limitations of their rule-based predecessors. LLM-based agents can provide tailored responses by adapting to diverse user contexts and incorporating prior interactions \cite{shukuri_meta-control_2023, 10.1145/3544548.3581503}. Furthermore, recent studies have demonstrated LLMs' ability to simulate specific characters \cite{cheong2022role, shao-etal-2023-character}, unlocking new possibilities for dynamic and personalized interventions in self-guided contexts. As a result, these capabilities have been successfully leveraged to enhance self-guided interventions across various domains. In the mental health domain, LLMs have been used to offer personalized support for self-led journaling \cite{10.1145/3613905.3650767} and to facilitate Cognitive Behavioral Therapy (CBT) \cite{na2024cbtllmchineselargelanguage}.

Building on these advancements, our study intends to harness the capabilities of LLM-based conversational agents to create dynamic and adaptive self-guided interventions for young adults' career exploration. Specifically, we integrated LLM agents simulating individuals' future selves into the letter-exchange exercise, aligning with the exercise’s core concept of fostering interaction between one’s present and future selves. By introducing a `\textbf{future-self agent},’ we sought to support the cognitively demanding aspects of the self-guided exercise by providing personalized guidance and facilitating dynamic, meaningful interactions with one’s future self.

However, prior studies have raised concerns about the potential drawbacks of over-reliance on AI, particularly its detrimental impact on critical thinking and self-reflection \cite{darwin2024critical, zhai2024effects}. In writing-based interventions like the letter-exchange exercise, where deep introspection and active participation are essential, excessive reliance on AI could undermine meaningful engagement in self-exploration and career development. These concerns, coupled with the potential benefits of AI in providing personalized support, underscore the need for a careful examination of how LLM agents influence young adults' engagement in the letter-exchange exercise. Through this exploration, we aimed to identify ways to balance the benefits of AI-driven support with the need to foster reflective and participatory engagement in self-guided interventions.

\begin{figure*}[!ht]
    \centering
    \includegraphics[width=440pt]{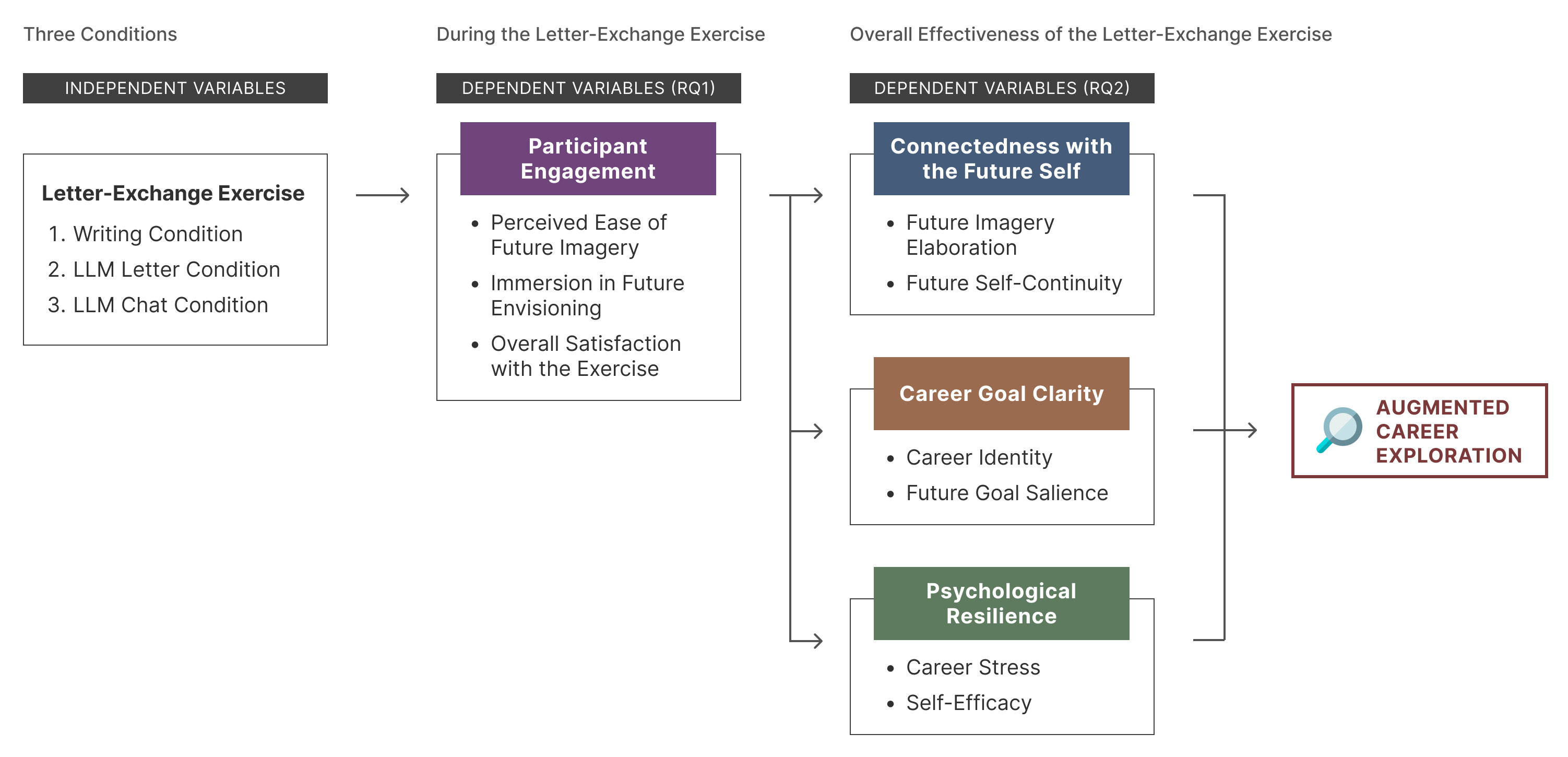}
    \caption{Research model overview}
    \Description{Figure 2 illustrates the overview of a research study model involving three conditions: Writing Condition, LLM Letter Condition, and LLM Chat Condition, categorized under `Independent Variables.' In the `During the Letter-Exchange Exercise' phase, `Dependent Variables (RQ1)' are displayed: Perceived Ease of Future Imagery, Immersion in Future Envisioning, and Overall Satisfaction with the Exercise. The `Overall Effectiveness after the Letter-Exchange Exercise' phase features `Dependent Variables (RQ2)': Connectedness with the Future Self (Future Imagery Elaboration and Future Self-Continuity), Career Goal Clarity (Career Identity and Future Goal Salience), and Psychological Resilience (Career Stress and Self-Efficacy). These lead to the outcome of Augmented Career Exploration.}
    \label{fig:2}
\end{figure*}

\section{Methods}

To investigate the potential of AI-augmented career exploration, we integrated LLM agents simulating participants' future selves into the conventional letter-exchange exercise and conducted a one-week between-subjects experiment (N=36). Participants completed the letter-exchange exercise, which was composed of (1) the \textbf{Send Session}, where all participants manually wrote letters to their future selves, and (2) the \textbf{Reply Session}, where they were randomly assigned to one of the following conditions:

\begin{itemize}
    \item \textbf{Writing Condition (N=12)}: Participants manually wrote replies to themselves from the perspective of their future selves.
    \item \textbf{LLM Letter Condition (N=12)}: Future-self agents generated replies from the perspective of participants’ future selves.
    \item \textbf{LLM Chat Condition (N=12)}: Participants engaged in chat conversations with their future-self agents.
\end{itemize}

The \textbf{Writing Condition} served as our baseline, replicating the conventional method proposed by Chishima and colleagues \cite{chishima2021conversation}. To examine the effect of future-self agents while maintaining the exercise’s original format, we included the \textbf{LLM Letter Condition}, where LLM agents generated responses from the perspective of participants' future selves. This condition aimed to preserve the reflective nature of letter writing while integrating future self agents. In contrast, the \textbf{LLM Chat Condition} was designed to facilitate a more dynamic and interactive approach by enabling participants to chat with their future self agent, reflecting the popularity of chat-based interactions in self-guided interventions \cite{song2024typingcureexperienceslarge}. By comparing these two LLM-based conditions, we sought to understand how differences in interaction style—static written responses versus real-time conversational exchanges—affect participants’ engagement and introspection during the exercise.

We employed a mixed-method approach to comprehensively evaluate the impact of the LLM-augmented letter-exchange exercise (see Figure \ref{fig:2}). First, we collected quantitative data on participants’ engagement during the exercise \textbf{(RQ1)} by evaluating how easily they could imagine and immerse themselves in future scenarios, as well as their overall satisfaction with the exercise. To assess the exercise's effectiveness \textbf{(RQ2)}, we quantitatively measured participants' sense of connectedness with their future selves, clarity of career goals, and psychological resilience at three time points (i.e., before, immediately after, and one week following the exercise). Finally, we collected qualitative data on participants’ experiences across the three conditions through interviews \textbf{(RQ3)}.

\subsection{Participants}

We recruited participants who reported being actively engaged in career exploration. A total of 36 participants (17 male, 19 female, average age=23.06 (\textit{SD}=1.84, Min=20, Max=27)) were recruited from a large research university in South Korea via announcements on university community websites. Participants were either (a) undergraduates expected to graduate within three years or (b) recent graduates not affiliated with any educational institutions or workplaces and engaged in career exploration and job searching. All participants were over the age of 19.

\subsection{Procedure}

\begin{figure*}[!ht]
    \centering
    \includegraphics[width=480pt]{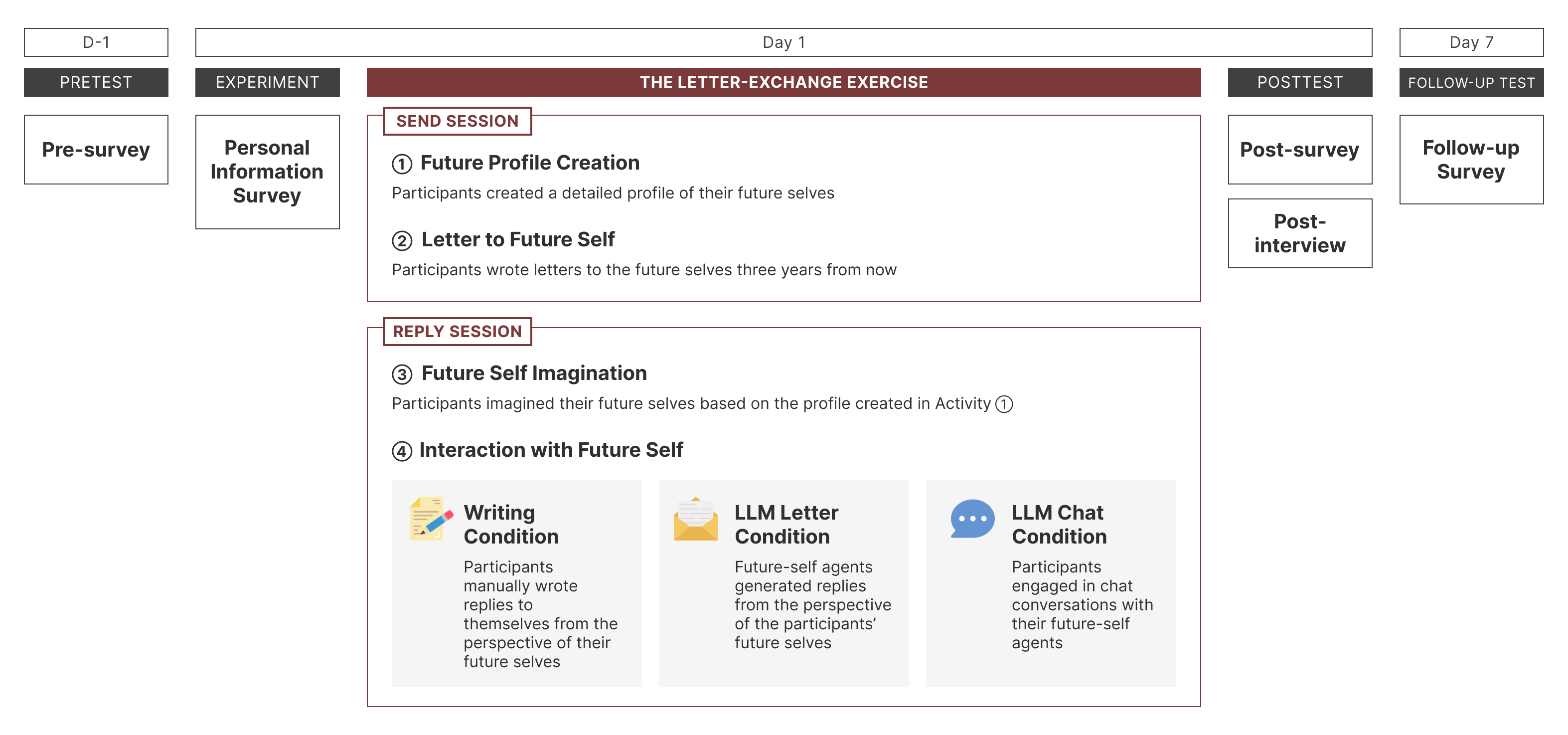}
    \caption{Study procedure}
    \Description{Figure 3 represents the procedure of the research study. It starts with a Pretest (D-1) involving a pre-survey. On Day 1, the experiment begins with a personal information survey. Then, participants perform `The Letter-Exchange Exercise', split into two sessions. First, the `Send Session' includes two activities: (1) Future Profile Creation, where participants create a detailed profile of their future selves, and (2) Letter to Future Self, where participants write letters to their future selves three years ahead. Then, the `Reply Session' involves two activities: (3) Future Self Imagination, where participants imagine their future selves based on the profile created in Activity 1, and (4) Interaction with Future Self, which varies based on the condition. The three conditions are Writing Condition (participants manually reply to themselves from the perspective of their future selves), LLM Letter Condition (future-self agents generate replies from the perspective of the participants' future selves), and LLM Chat Condition (participants engage in chat conversations with their future-self agents). Following the experiment, a Posttest includes a post-survey and a post-interview on Day 1. On Day 7, a Follow-up Test involves a follow-up survey.}
    \label{fig:3}
\end{figure*}

We conducted the experiment from May to July 2024. The whole procedure spanned over one week and included four stages: (1) pretest, (2) in-person experiment (letter-exchange exercise), (3) posttest, and (4) follow-up test. All study materials and procedures received approval from the Institutional Review Board (IRB) of the university where the study was hosted, and all procedures and interview recordings were conducted with the informed consent of the participants. The entire experiment, including the study materials and analysis, was conducted in Korean. During the paper-writing process, participants’ quotes were translated into English by bilingual Korean-English speakers in the research team.

\subsubsection{Pretest}

\label{sec:pretest}
A day before the in-person experiment, we asked participants to complete an online pre-survey, which took approximately five minutes. The pre-survey consisted of two main components: (1) measures evaluating (a) Connectedness with the Future Self, (b) Career Goal Clarity, and (c) Psychological Resilience as a baseline for \textbf{RQ2} and (2) measures of participants' current career development status, including measures of career exploration \cite{stumpf1983development} and career planning \cite{gould1979characteristics}. Given the demographic characteristics of our participant group, who were at a stage of career exploration and planning, we included these measures to assess their current career exploration context. These components provided key information for tailoring the LLM agents to each participant's unique career development context.

\subsubsection{In-person Experiment: The Letter-Exchange Exercise}

\label{sec:experiment}
We conducted the in-person experiment in a quiet room with two researchers per session. Before proceeding with the experiment, we thoroughly informed the participant about the study's purpose and procedures and obtained their consent. The session then proceeded as follows.

\begin{figure*}[!ht]
    \centering
    \includegraphics[width=280pt]{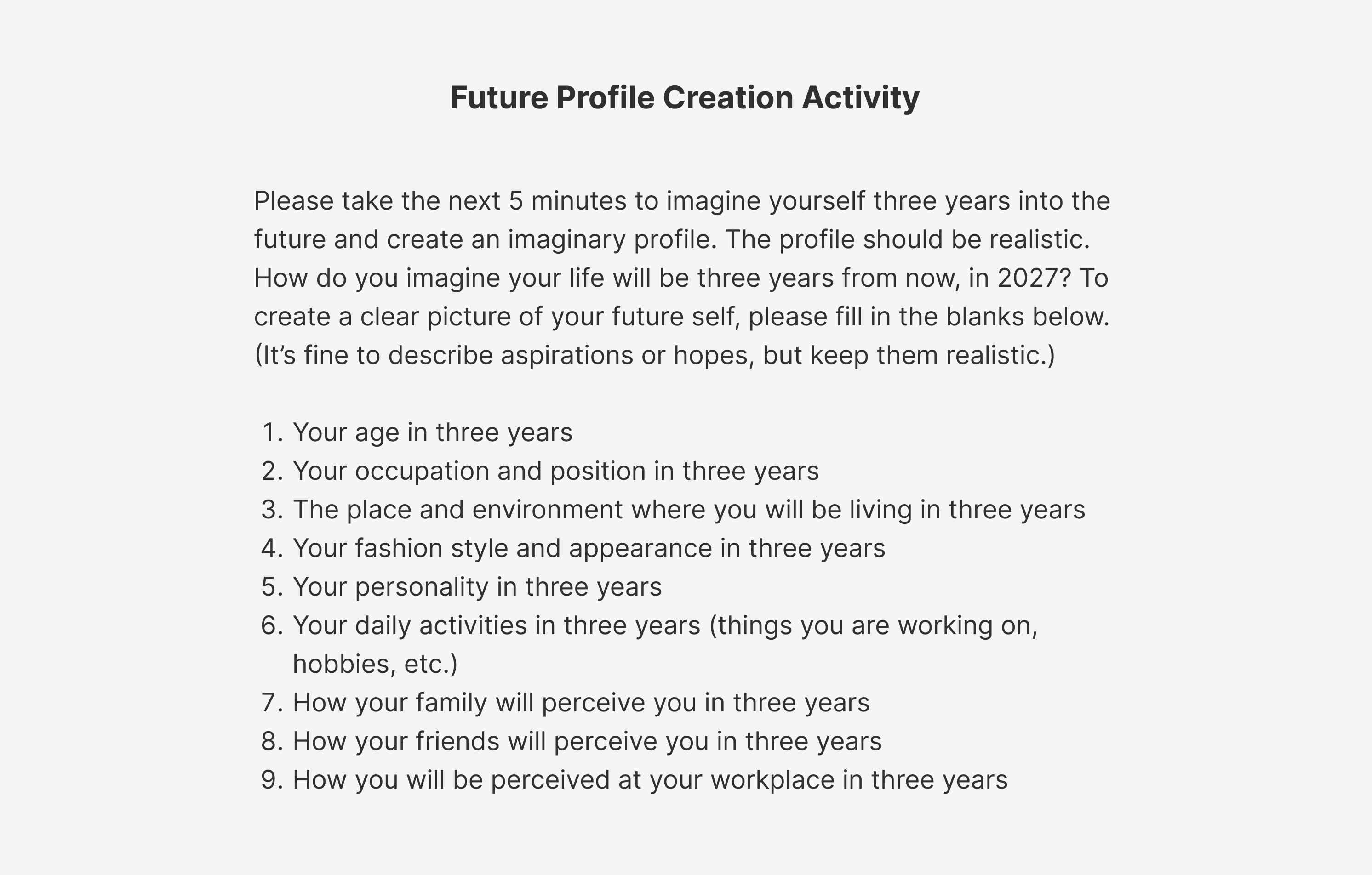}
    \caption{Template for future profile (3 years later)}
    \Description{Figure 4 displays instructions for the `Future Profile Creation Activity.' Participants are asked to take 5 minutes to imagine themselves three years into the future and create a realistic profile based on their life in 2027. The instructions emphasize that it's acceptable to describe aspirations or hopes, but they should remain realistic. Below the instructions is a list of nine prompts to guide participants in building their future profile: Your age in three years / Your occupation and position in three years / The place and environment where you will be living in three years / Your fashion style and appearance in three years / Your personality in three years / Your daily activities in three years (things you are working on, hobbies, etc.) / How your family will perceive you in three years / How your friends will perceive you in three years / How you will be perceived at your workplace in three years.}
    \label{fig:4}
\end{figure*}

\begin{figure*}[!ht]
    \centering
    \includegraphics[width=420pt]{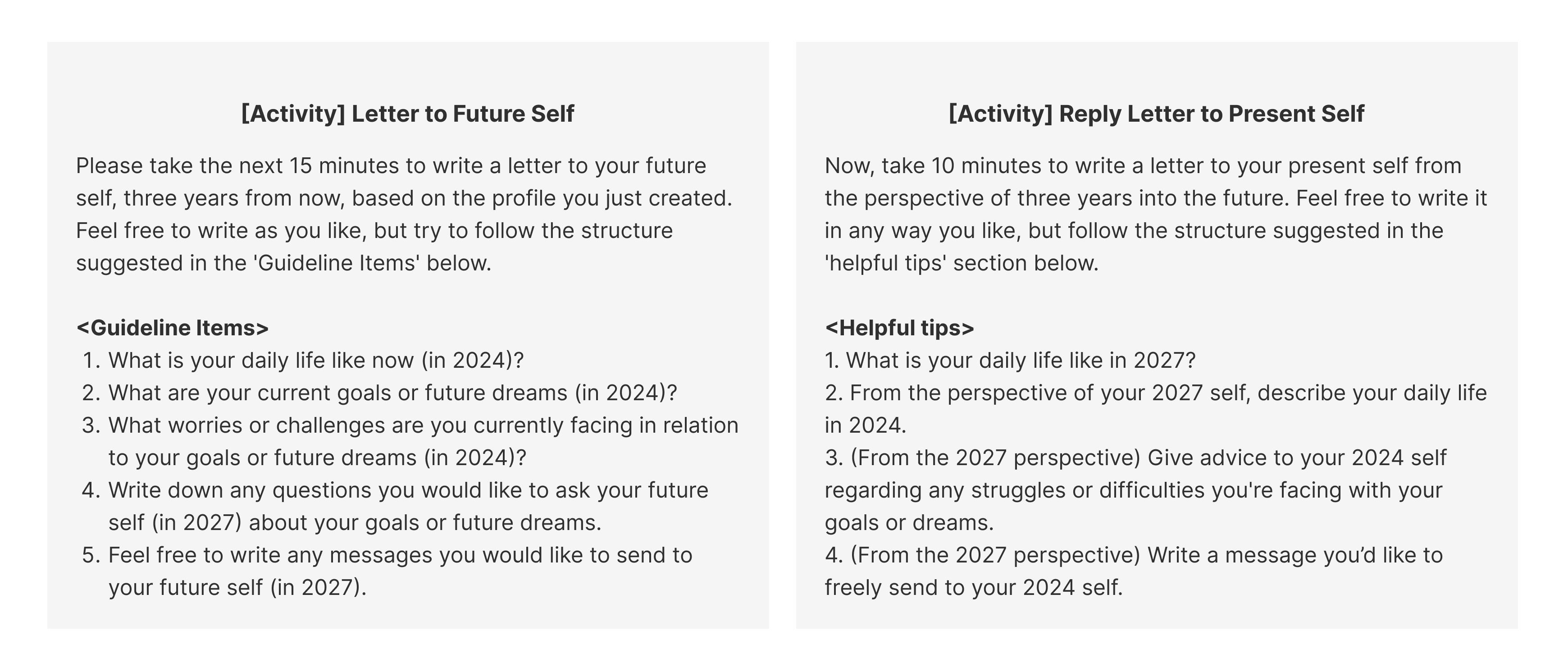}
    \caption{Instructions for two activities: (1) writing a letter to future self and (2) writing a reply to present self}
    \Description{Figure 5 contains two sections, each providing activity instructions. The first section on the left side is about the `Letter to Future Self' activity. Participants are given 15 minutes to write a letter to their future self three years ahead, based on a created profile. The letter should follow the structure provided in five `Guideline Items': What is your daily life like now (in 2024)? / What are your current goals or future dreams (in 2024)? / What worries or challenges are you currently facing in relation to your goals or future dreams (in 2024)? / Write down any questions you would like to ask your future self (in 2027) about your goals or future dreams. / Feel free to write any messages you would like to send to your future self (in 2027). The second section on the right side is about the `Reply to Present Self' activity. Participants are given 10 minutes to write a reply to their present self from the perspective of their future self three years into the future. The structure is suggested in four `Helpful tips': What is your daily life like in 2027? / From the perspective of your 2027 self, describe your daily life in 2024. / (From the 2027 perspective) Give advice to your 2024 self regarding any struggles or difficulties you're facing with your goals or dreams. / (From the 2027 perspective) Write a message you’d like to freely send to your 2024 self.}
    \label{fig:5}
\end{figure*}

\begin{enumerate}
    \item \textbf{Personal Information Survey (5 min)}: Participants completed an online survey that gathered demographic information, assessed personality traits using the BFI-2-S survey items \cite{soto2017short}, and measured personal values using the PVQ scale \cite{lindeman2005measuring}. These data were utilized to comprehensively incorporate personal traits in the implementation of future-self agents.
    
    \item \textbf{Send Session (20 min)}: The Send Session was conducted the same in all conditions.
    \begin{itemize}
        \item \textbf{Future Profile Creation Activity (5 min)}: Participants were instructed to complete a detailed profile of themselves three years ahead using a provided template (see Figure \ref{fig:4}). They were guided to ensure that the profile reflected realistic future aspirations. The three-year time frame was selected as it provides a temporal distance sufficient to capture meaningful career development progress while remaining close enough to allow for vivid and relatable imagery \cite{chishima2021conversation}.
        \item \textbf{Letter to Future Self Activity (15 min)}: Based on created profiles, participants were asked to write a letter to their future selves three years from now. Guidelines were provided to facilitate the writing process (see Figure \ref{fig:5}-Left).
    \end{itemize}
    
    \item \textbf{Reply Session (20 min)}: After a 5-minute break, the Reply Session began.
    \begin{itemize}
        \item \textbf{Future Self Imagination Activity (5 min)}: Participants were asked to imagine themselves in three years, living the life described in their profiles.
        \item \textbf{Interaction with Future Self Activity (15 min)}: The final activity of the letter-exchange exercise varied depending on the condition:
        \begin{itemize}
            \begin{sloppypar}
            \item \textbf{Writing Condition}: Following the conventional self-guided method, participants wrote a reply to their present selves from the perspective of their future selves. Guidelines were provided to help them compose replies (see Figure \ref{fig:5}-Right).
            \end{sloppypar}
            \item \textbf{LLM Letter Condition}: Participants received a reply generated by future-self agents, presented as a letter from their future selves. The LLM agents generated the reply based on the letter participants wrote in the Send Session and adhered to the same guidelines provided in the Writing Condition.
            \item \textbf{LLM Chat Condition:} Participants engaged in a real-time chat with future-self agents for about 15 minutes, simulating a conversation with their future selves. They were encouraged to discuss career-related concerns or challenges with suggested topics, which were comparable to those in the LLM Letter Condition, to guide the conversation.
        \end{itemize}
    \end{itemize}
\end{enumerate}

\subsubsection{Posttest and Interview}

After completing the exercise, participants filled out a post-survey, which took approximately 5 minutes. The post-survey comprised two components: (1) measures evaluating participants' engagement during the letter-exchange exercise \textbf{(RQ1)} and (2) the same effectiveness measures \textbf{(RQ2)} from the pre-survey (see Section \ref{sec:pretest}). Following the survey, we conducted a semi-structured post-interview to gain deeper insights into participants' experiences \textbf{(RQ3)}.

\subsubsection{Follow-Up Test}

One week after the experiment, participants completed a follow-up survey online. The one-week interval was chosen to evaluate the lasting impact of the intervention beyond its immediate post-exercise outcomes while minimizing potential confounding factors that could emerge over longer periods. The follow-up survey comprised the same measures from the pre- and post- surveys to evaluate the effectiveness of the exercise \textbf{(RQ2)}. Upon completion of the follow-up survey, participants were compensated with 30,000 Korean Won (equivalent to \$23 USD).

\subsection{Technical Implementation of the LLM-based Agent Simulating Participants’ Future Selves}

In developing personalized future-self agents using LLMs, we took a comprehensive approach to simulate realistic interactions with participants. This section outlines the process of creating future-self agents, detailing the knowledge structure and the implementation of different interaction modalities.

\subsubsection{Creating Future-Self Agents}

\begin{figure*}[!ht]
    \centering
    \includegraphics[width=500pt]{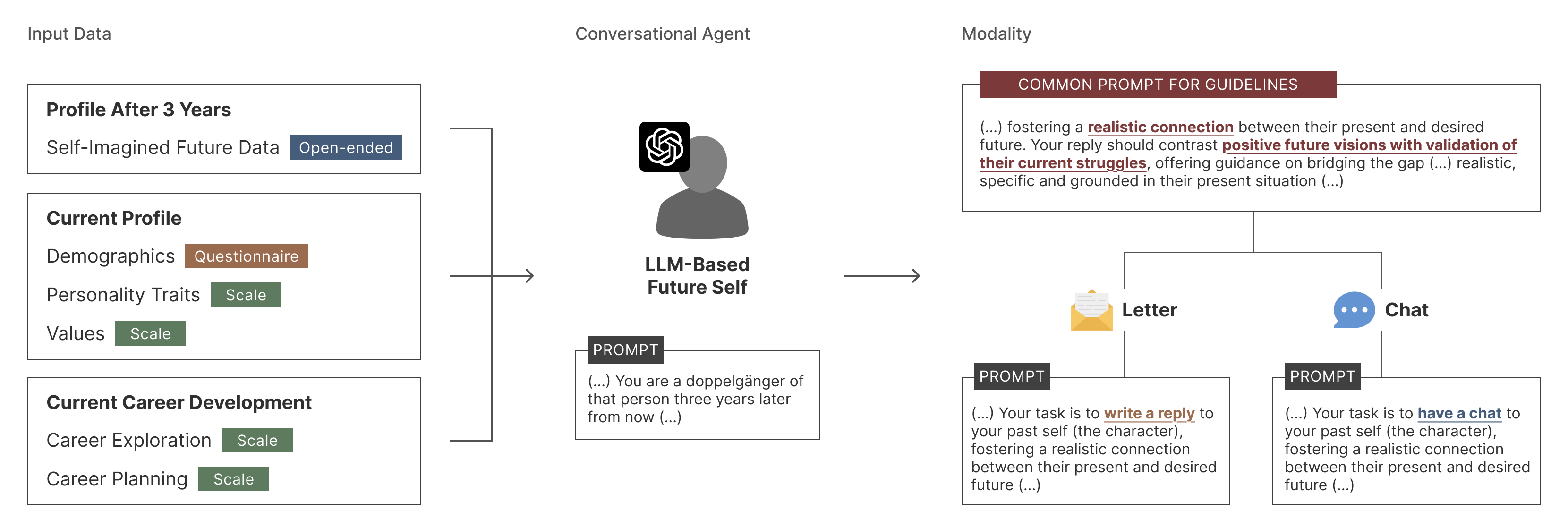}
    \caption{Future-self LLM agent system overview}
    \Description{A flowchart illustrates the system of the future-self agent. The diagram shows three main components. On the far left side, the `Input Data' component includes (1) `Profile after 3 years' (self-imagined future data from an open-ended survey), (2) `Current Profile' (demographics from questionnaire, personality traits and values based on scales), and (3) `Current Career Development' (career exploration and planning based on scales). This component connects to the next component in the middle. In the middle, the `Conversational Agent' component illustrates an LLM-based future-self agent represented by a ChatGPT logo and a person icon. The agent is prompted to act as a doppelgänger of the person three years in the future. This component connects to the next component on the far right side. On the far right side, the `Modality' component includes two interaction methods - `Letter' and `Chat'. Each shows example prompts that instruct the agent to write a reply or have a chat with the present self, fostering a connection between present and desired future. Additionally, a common prompt for both modalities is shown, instructing the agents to contrast positive future visions with current struggles and offer guidance on bridging the gap between the present and future through self-regulation and goal pursuit.}
    \label{fig:6}
\end{figure*}

\begin{sloppypar}
To simulate realistic future selves, we structured the agent's knowledge into three key components (see Figure \ref{fig:6}). One of the core components was the `Profile After 3 Years' that each participant created, which served as the foundation for future selves embodied by LLM agents. 
\end{sloppypar}

While simulating one's future self was our primary focus, we recognized the importance of grounding this simulation in the participant's present circumstances. Thus, we included a `Current Profile' component, which provided essential background for connecting the participant's present with their desired future. Building upon the SPeCtrum framework for representing one's multidimensional identity into LLM-based agents \cite{lee2025spectrumgroundedframeworkmultidimensional}, we incorporated participants' demographic information directly into the prompt, along with personality traits and values, which were converted into natural language from the Big Five Inventory-2-Short Form (BFI-2-S) \cite{soto2017short} and Portrait Values Questionnaire (PVQ) \cite{schwartz2009basic} scores. To further refine this information, we applied the Chain of Density (CoD) prompting technique for progressive summarization \cite{serapio2023personality}, generating a comprehensive summary of one's personality traits and value systems.

The third component, `Current Career Development Status,' accounted for each participant's unique stage in career preparation. This incorporated measures of career exploration \cite{stumpf1983development} and career planning \cite{gould1979characteristics} that we collected at the pretest stage. We converted these scales into natural language statements to provide context for the agent's interactions. A comprehensive overview of the knowledge structure used to create the future-self agent can be found in Appendix \ref{A.1}.

To foster an authentic experience of communicating with one's future self, we developed specific prompting strategies (see Appendix \ref{A.2}). First, we explicitly defined the agent's role with the instruction: \textit{"You are a doppelgänger of that person three years later from now."} Then, to ensure consistency and accuracy in the agent’s responses, we provided guidelines for knowledge integration: \textit{"Using the provided profile (knowledge), replicate the person's attitudes, thoughts, and mannerisms three years from now as accurately as possible."} Further, we ensured that the agent's portrayal of the future was positive yet realistic by instructing it to \textit{"describe daily life three years in the future that aligns with their goals, keeping it realistic and grounded in their present situation"} and \textit{"Your reply should contrast positive future visions with validation of their current struggles."} This approach aimed to motivate participants by encouraging introspection without overly idealizing or negatively framing their future.

\subsubsection{Designing the Interaction for Each Modality}

To augment the interaction with future-self agents, we implemented two interaction modalities: letter-based and chat-based. For letter-based interactions, we utilized OpenAI's GPT-4o model, the most advanced model available at the time of the study. For chat-based interactions, we developed custom versions of ChatGPT using GPTs.

In two LLM conditions, the process remained consistent; participants wrote a letter to their future selves, and the agent responded according to the specific modality in each condition. To ensure comparability across all conditions (i.e., Writing, LLM Letter, and LLM Chat), we standardized the overall structure of the future self's responses to align with the Writing Condition. The guidelines provided to participants in the Writing Condition were included in the input prompts for all agents, regardless of modality, ensuring that key topics were addressed uniformly across all conditions. We also implemented strategies to handle the potential limitations of LLMs. For instance, to avoid generic and irrelevant responses, we explicitly instructed the agent: \textit{"Even if the person mentions something not specific in the profile, use your best guesses and imagination to respond."} Additionally, to minimize perceptual dissonance in communication style, the agent was instructed to adjust its tone and conversational style based on the provided profile information, such as age and personality traits.

Further, we customized the prompts to suit each modality for more natural exchanges. For letter-based interactions, the agent was prompted to respond to the present self in approximately 200-word letters, maintaining consistency in the length and depth of responses with the Writing Condition (see Appendix \ref{A.2}). For chat-based interactions, we instructed the agent to engage in a chat with their present self, limiting each message to three sentences or fewer to simulate a realistic chat experience. To maintain engagement and encourage dynamic conversation, the prompt included the instruction: \textit{"Ask a question at least once every three exchanges."} To prevent off-topic divergence, which was a potential issue in the open-ended chat format, we also included the direction: \textit{"If the conversation goes off-topic, steer it back to the main topic."} (see Appendix \ref{A.3}).

\subsection{Quantitative Measures}

\subsubsection{Participants’ Engagement During the Letter-Exchange Exercise (RQ1)}

To assess participants' engagement with the exercise \textbf{(RQ1)}, we measured three variables immediately after the exercise. We selected these variables, considering that the letter-exchange exercise was designed to elicit concrete future imagery and foster immersion in it. Each item was measured on a 7-point Likert scale.

\begin{itemize}
    \item \textbf{Perceived Ease of Future Imagery during the Exercise.} Participants responded to the statement, \textit{`During the activity, I found it easy to elicit images or sensory feelings about my future.'}
    \item \textbf{Immersion in Future Envisioning during the Exercise.} Participants rated their agreement with the statement, \textit{`During the activity, I felt deeply engaged with the future I was describing.'}
    \item \textbf{Overall Satisfaction with the Exercise.} Participants responded to two statements, and their ratings were averaged: \textit{`Overall, I am satisfied with the activity' }and\textit{ `I am generally satisfied with the content produced through the activity.'} (Cronbach's alpha = .827)
\end{itemize}

\subsubsection{Effectiveness of the Three Types of Letter-Exchange Exercise (RQ2)}

To assess exercise effectiveness \textbf{(RQ2)}, we measured the following variables across three time points (i.e., pre-, post-, and follow-up). These variables were selected to align with the initial objectives of the letter-exchange exercise \cite{chishima2021conversation} to enhance future orientation, career development, and psychological support. Each of these constructs consisted of multiple sub-scales.

\begin{itemize}
    \item \textbf{Connectedness with the Future Self.} Participants' perceived connection to their future selves was measured based on (1) vividness with which they envisioned their future and (2) their sense of connection to that future.
    \begin{itemize}
        \item \textit{Future Imagery Elaboration.} We used a subset of the Future Work Self Scale \cite{strauss2012future} (e.g., \textit{`The image of my future is very vivid'}). Five items were measured on a 7-point Likert scale and were averaged (Cronbach's alpha = .912).
        \begin{sloppypar}
        \item \textit{Future Self-Continuity.} We employed the Future Self-Continuity Scale \cite{ersner2009don} based on the Inclusion of Other in the Self Scale (IOS) \cite{aron1992inclusion}. Participants reported how connected they felt to their future selves using a single item on a 7-point Likert scale.
        \end{sloppypar}
    \end{itemize}
    
    \item \textbf{Career Goal Clarity.} The level of participants' career development was assessed based on (1) self-awareness of their career identities and (2) clarity of their career goals. Each item was measured on a 7-point Likert scale.
    \begin{itemize}
        \item \textit{Career Identity.} We adapted the 4-item Career Identity Scale \cite{dobrow2005developmental} (e.g., \textit{`I have established a clear career identity'}). An exploratory factor analysis was conducted to eliminate one item that did not meet the factor loading threshold of .60, after which the remaining items were averaged (Cronbach's alpha = .873).
        \item \textit{Future Goal Salience.} We utilized a subset of the Future Work Self Scale \cite{strauss2012future}, using two items (e.g., \textit{`I am very clear about who I want to become in the future'}) (Cronbach's alpha = .897).
    \end{itemize}
    
    \item \textbf{Psychological Resilience.} Participants' career resilience and motivation were evaluated based on (1) their career-related stress level and (2) their overall self-efficacy. Each item was measured on a 7-point Likert scale.
    \begin{itemize}
        \item \textit{Career Stress.} We used five items from the Career Stress Scale \cite{choi2011development} to assess stress related to career uncertainty. Participants rated statements such as \textit{`I feel frustrated because I doubt that my career plans will work out'} (Cronbach's alpha = .868).
        \item \textit{Self-Efficacy.} We employed the 8-item General Self-Efficacy Scale \cite{chen2001validation} with participants responding to statements such as \textit{`I can achieve most of the goals that I have set for myself'} (Cronbach's alpha = .933).
    \end{itemize}
\end{itemize}

\subsection{Interview Protocols (RQ3)}

We conducted semi-structured interviews to gain deeper insights into participants' experiences across three conditions. The interviews began with general questions about participants’ career development progress and concerns, followed by their experiences with the letter-exchange exercise. This included discussions on its impact on their career goals, future imagery, psychological motivation, and any challenges they encountered during the exercise.

Condition-specific questions were then introduced. For the Writing Condition, we explored how composing a letter to their future selves and writing a reply from the future perspective influenced their sense of connectedness and relationship with their future selves. In the LLM Letter and Chat Conditions, participants were asked to reflect on their impressions of the LLM-generated letters or chat interactions. We further examined participants’ emotional responses to interacting with future-self agents, as well as their perceptions of the content (e.g., \textit{`Was the depicted future realistic?'}) and format (e.g., \textit{`Did the agent's tone align with your usual communication style?'}) of the interactions.

\subsection{Data Analysis}

\subsubsection{Quantitative Data Analysis}

We analyzed the quantitative data using the Jamovi software \cite{csahin2019jamovi}, applying appropriate statistical methods based on the nature of the data for each RQ. To examine the differences in participants' engagement across the three types of letter-exchange exercise \textbf{(RQ1)}, we conducted a One-way ANOVA. Games-Howell post-hoc tests were employed for pairwise comparisons to account for unequal variances across conditions. To evaluate the effectiveness of the exercises across three time points \textbf{(RQ2)}, we used a Mixed Model ANOVA. The between-subjects factor was the exercise condition (i.e., Writing, LLM Letter, and LLM Chat), while the within-subjects factor was the time point (i.e., pre-, post-, and follow-up). Bonferroni corrections were applied to all pairwise comparisons.

\subsubsection{Qualitative Data Analysis}

For \textbf{RQ3} (Participants' Experience), we analyzed the interview data using a thematic analysis method \cite{braun2012thematic}. First, the post-interviews were fully transcribed in Korean by authors who facilitated the experiments. Then, four authors of the research team individually coded the entire interview transcripts using Atlas.Ti software \cite{woods2016advancing}, which resulted in 337 initial codes. Following this, we carried out an iterative process of reviewing the codes, discussing interpretations, and grouping the codes into similar categories. This process resulted in seven main themes, including \textit{Perception of Future Self}, \textit{Career Exploration Patterns in Each Condition}, \textit{Advantages of LLM Integration}, and \textit{Effects of the Exercise}. Finally, we created affinity diagrams to further examine the relationships across these main themes. After completing the analysis, the qualitative data were translated into English by bilingual authors.


\begin{figure*}[!ht]
    \centering
    \includegraphics[width=480pt]{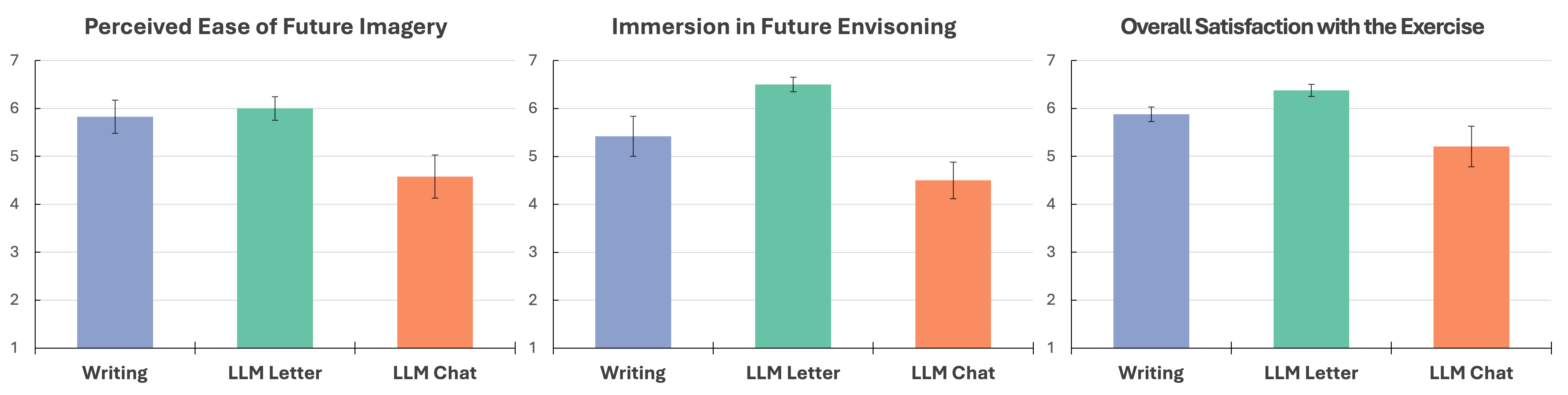}
    \caption{Quantitative results for Perceived Ease of Future Imagery, Immersion in Future Envisioning, and Overall Satisfaction with the Exercise. The error bars represent standard errors. }
    \Description{Figure 7 presents three bar graphs, visualizing mean scores for three variables, with 95 percent confidence intervals. First bar graph shows the quantitative results for the `Perceived Ease of Future Imagery during the Exercise' measure. The second bar graph shows the quantitative results for the `Immersion in Future Envisioning during the Exercise' measure. The last bar graph illustrates the quantitative results for the `Overall Satisfaction with the Exercise' measure. In each bar graph, three conditions are compared: Writing, LLM Letter, and LLM Chat.}
    \label{fig:7}
    \vspace{-0.1cm}
\end{figure*}

\section{Quantitative Findings (RQ1, RQ2)}

In this section, we report the quantitative results for \textbf{RQ1} (Participants’ Engagement) and \textbf{RQ2} (Exercise Effectiveness).

\subsection{Participants’ Engagement During the Three Types of Letter-Exchange Exercise (RQ1)}

To examine participant engagement with the exercise \textbf{(RQ1)}, we measured three variables immediately after the exercise (post-): (1) Perceived Ease of Future Imagery during the Exercise, (2) Immersion in Future Envisioning during the Exercise, and (3) Overall Satisfaction with the Exercise. We ran a one-way ANOVA to analyze differences in participant engagement measures across the LLM conditions (i.e., LLM Letter, LLM Chat) and the conventional method (i.e., Writing). 

\subsubsection{Perceived Ease of Future Imagery during the Exercise}

We found a statistically significant effect of condition on perceived ease of future imagery during the exercise (F(2, 20.8) = 3.75, p = .041, $\eta^2$= .26). The Games-Howell post-hoc test indicated that participants in the LLM Letter Condition (\textit{M} = 6.00, \textit{SD} = .85) reported significantly higher perceived ease of eliciting future imagery compared to those in the LLM Chat Condition (\textit{M} = 4.58, \textit{SD} = 1.56) (p = .034; see Figure \ref{fig:7}). No significant differences were observed between the Writing (\textit{M} = 5.83, \textit{SD} = 1.19) and the LLM Letter Condition (p = .91), nor between the Writing and the LLM Chat Condition (p = .09). These findings suggest that participants found it easier to envision their future selves through LLM-generated letters than through chat-based interactions while reporting comparable ease of future imagery between the LLM Letter and the Writing Condition.

\subsubsection{Immersion in Future Envisioning during the Exercise}

We found a statistically significant effect of condition on perceived immersion during the exercise (F(2, 18.0) = 13.2, p < .001, $\eta^2$= .59). The post-hoc test indicated that participants in the LLM Letter Condition exhibited significantly higher perceived immersion compared to those in the LLM Chat Condition (\textit{M} = 4.50, \textit{SD} = 1.31) (p < .001). Additionally, participants in the LLM Letter Condition (\textit{M} = 6.50, \textit{SD} = .52) reported marginally higher perceived immersion than those in the Writing Condition (\textit{M} = 5.42, \textit{SD} = 1.44) (p = .06; see Figure \ref{fig:7}). No significant difference was found between the Writing and the LLM Chat Condition (p = .25). This demonstrated that participants experienced deeper immersion in future envisioning with the LLM letters compared to chat-based interactions while showing marginally higher levels of immersion with the LLM Letter than the Writing Condition.

\subsubsection{Overall Satisfaction with the Exercise}

We found a statistically significant effect of condition on participants' satisfaction with the exercise and its output (e.g., the content of letters or chats) (F(2, 19.9) = 5.50, p < .05, $\eta^2$= .35). The post-hoc test showed that participants in the LLM Letter Condition (\textit{M} = 6.38, \textit{SD} = .43) reported significantly higher satisfaction than those in the Writing (\textit{M} = 5.88, \textit{SD} = .52) (p = .04) and the LLM Chat Condition (\textit{M} = 5.21, \textit{SD} = 1.46) (p = .05; see Figure \ref{fig:7}). No significant difference was found between the Writing and the LLM Chat Condition (p = .33). This suggests that participants exhibited higher levels of satisfaction with the LLM-generated letters compared to both the chat-based interactions and the conventional self-guided method.

\begin{figure*}[!ht]
    \centering
    \includegraphics[width=485pt]{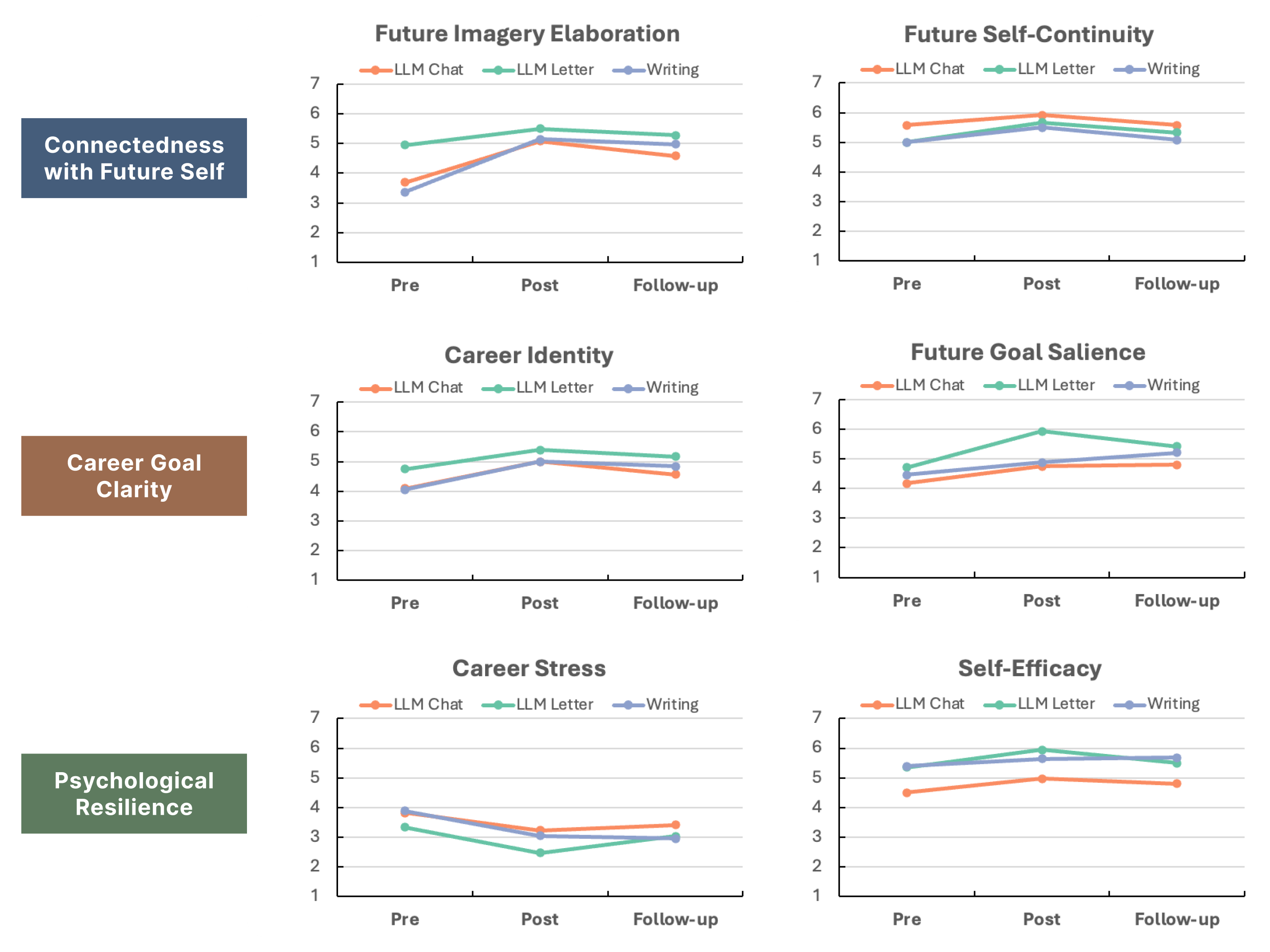}
    \caption{Trends in Connectedness with the Future Self (Top), Career Goal Clarity (Middle), Psychological Resilience (Bottom) sub-scales across three time points}
    \Description{Figure 8 presents three sets of two line graphs with plots side by side. The first set shows two graphs for Connectedness with Future Self. The left graph shows how the mean scores of Future Imagery Elaboration increase across three-time points (Pre, Post, and Follow-up) for three conditions: Writing, LLM Letter, and LLM Chat. The right graph shows how the mean scores of Future Self-Continuity fluctuate across the same three time points for the same three conditions. The second set shows two graphs for Career Goal Clarity. The left graph shows how the mean scores of Career Identity increase across three time points (Pre, Post, and Follow-up) for three conditions: Writing, LLM Letter, and LLM Chat. The right graph shows how the mean scores of Future Goal Salience increase across the same three time points for the same three conditions. The third set shows two graphs for Psychological Resilience. The left graph shows how the mean scores of Career Stress decrease across three time points (Pre, Post, and Follow-up) for three conditions: Writing, LLM Letter, and LLM Chat. The right graph shows how the mean scores of Self-Efficacy fluctuate across the same three time points for the same three conditions.}
    \label{fig:8}
\end{figure*}

\subsection{Effectiveness of the Letter-Exchange Exercise depending on Condition (RQ2)}

\begin{sloppypar}
To address \textbf{RQ2}, we evaluated the effectiveness of the letter-exchange exercise by measuring variables across three dimensions: Connectedness with the Future Self, Career Goal Clarity, and Psychological Resilience at three time points (i.e., pre-, post-, and follow-up). A 3 (time point) × 3 (condition) mixed ANOVA with Bonferroni-adjusted post hoc tests was conducted to examine differences across conditions and time points.
\end{sloppypar}

Overall, our analysis revealed that all three conditions were comparably effective in enhancing the measured dimensions, with all sub-measures showing immediate improvements after the exercise. However, while most measures sustained these increases at the one-week follow-up, two dependent variables—Future Self-Continuity and Self-Efficacy—did not maintain their gains across any condition. In general, these findings suggest that LLM-powered approaches could perform comparably to the conventional self-guided method.

\subsubsection{Connectedness with the Future Self}

We found a significant main effect of time for both Future Imagery Elaboration and Future Self-Continuity, with no significant main effect of condition for either measure.
\setlength{\parskip}{0pt}

\begin{itemize}
    \item \textit{Future Imagery Elaboration.} There was a significant main effect of time on Future Imagery Elaboration (F(2, 66) = 30.61, p < .001, $\eta^2$= .155). Bonferroni-adjusted post-hoc tests revealed that scores both immediately after the exercise (post-) (\textit{M} = 5.24, \textit{SD} = 0.98; p < .001) and at follow-up (\textit{M} = 4.94, \textit{SD} = 1.23; p < .001) were significantly higher than pre-exercise levels (\textit{M} = 4.00, \textit{SD} = 1.48). These results indicate that participants maintained enhanced Future Imagery Elaboration through the one-week follow-up period regardless of condition.
    
    \item \textit{Future Self-Continuity.} A significant main effect of time was observed for Future Self-Continuity (F(2, 66) = 4.473, p = .01, $\eta^2$= .034). Post-hoc tests revealed that post-exercise scores (\textit{M} = 5.69, \textit{SD} = 1.09; p = .01) were significantly higher than pre-exercise scores (\textit{M} = 5.19, \textit{SD} = 1.26). However, a decrease was observed at follow-up (\textit{M} = 5.33, \textit{SD} = 1.04; p = .04) compared to the post-exercise scores, though scores remained higher than pre-exercise. This suggests a positive short-term effect of the exercise while highlighting the need for additional interventions to maintain long-term effects.
\end{itemize}

\subsubsection{Career Goal Clarity}

\begin{sloppypar}
A Mixed ANOVA with Bonferroni-adjusted post-hoc tests was conducted to analyze differences across conditions and time points for Career Goal Clarity ratings. The analysis revealed significant main effects of time for both measures, with no significant main effect of condition. Both Career Identity and Future Goal Salience showed significant improvements that were maintained through the follow-up period across all conditions.
\end{sloppypar}

\begin{itemize}
    \item \textit{Career Identity.} We found a significant main effect of time for Career Identity (F(2, 66) = 16.19, p < .001, $\eta^2$= .067). Post-hoc tests indicated that scores both immediately after the exercise (post-) (\textit{M} = 5.13, \textit{SD} = 1.12; p < .001) and at follow-up (\textit{M} = 4.85, \textit{SD} = 1.35; p = .002) were significantly higher than pre-exercise scores (\textit{M} = 4.30, \textit{SD} = 1.44). This indicates that participants demonstrated improved Career Identity immediately after the exercise, maintaining this enhancement through the one-week follow-up period regardless of condition.

    \item \textit{Future Goal Salience.} A significant main effect of time was found for Future Goal Salience (F(2, 66) = 10.55, p < .001, $\eta^2$= .051). Post-hoc tests indicated that scores both immediately after the exercise (post-) (\textit{M} = 5.18, \textit{SD} = 1.33; p = .001) and at follow-up (\textit{M} = 5.14, \textit{SD} = 1.41; p = .003) were significantly higher than pre-exercise scores (\textit{M} = 4.44, \textit{SD} = 1.66). This shows that the exercises enhanced participants' clarity of future goals, with improvements persisting through the follow-up period across all conditions.
\end{itemize}

\subsubsection{Psychological Resilience}

A Mixed ANOVA with Bonferroni-adjusted post-hoc tests was conducted to analyze differences across conditions and time points for sub-measures in Psychological Resilience. The analysis showed significant main effects of time for both measures, with no significant main effect of condition. Career Stress showed sustained improvements through the follow-up period, while Self-Efficacy demonstrated only immediate enhancement that was not maintained at follow-up.

\begin{itemize}
    \item \textit{Career Stress.} We found a significant main effect of time for Career Stress (F(2, 66) = 19.88, p < .001, $\eta^2$= .061). Post-hoc tests revealed that scores both immediately after the exercise (post-) (\textit{M} = 2.92, \textit{SD} = 1.26; p < .001) and at the follow-up (\textit{M} = 3.13, \textit{SD} = 1.23; p < .001) were significantly lower than pre-exercise scores (\textit{M} = 3.68, \textit{SD} = 1.34). This indicates that participants experienced reduced Career Stress levels immediately after the exercise, and this alleviation was sustained through the one-week follow-up period across all conditions.

    \item \textit{Self-Efficacy.} A significant main effect of time was observed for Self-Efficacy (F(2, 66) = 10.31, p < .001, $\eta^2$= .028). Post-hoc tests revealed that scores immediately after the exercise (post-) (\textit{M} = 5.52, \textit{SD} = 1.03; p < .001) were significantly higher than scores before the exercise (pre-) (\textit{M} = 5.08, \textit{SD} = 1.17). However, at follow-up (\textit{M} = 5.33, \textit{SD} = 1.00), no significant difference was found compared to pre-exercise scores (p = .34). This suggests that while the exercises had an immediate positive impact on participants' self-efficacy, this effect was not fully sustained through the follow-up period across all conditions.
\end{itemize}


\section{Qualitative Findings (RQ3)}

To address RQ3, we analyzed interview data to explore how each condition shaped participants’ career exploration processes, revealing strengths and challenges. 

\begin{figure*}
    \centering
    \includegraphics[width=0.95\linewidth]{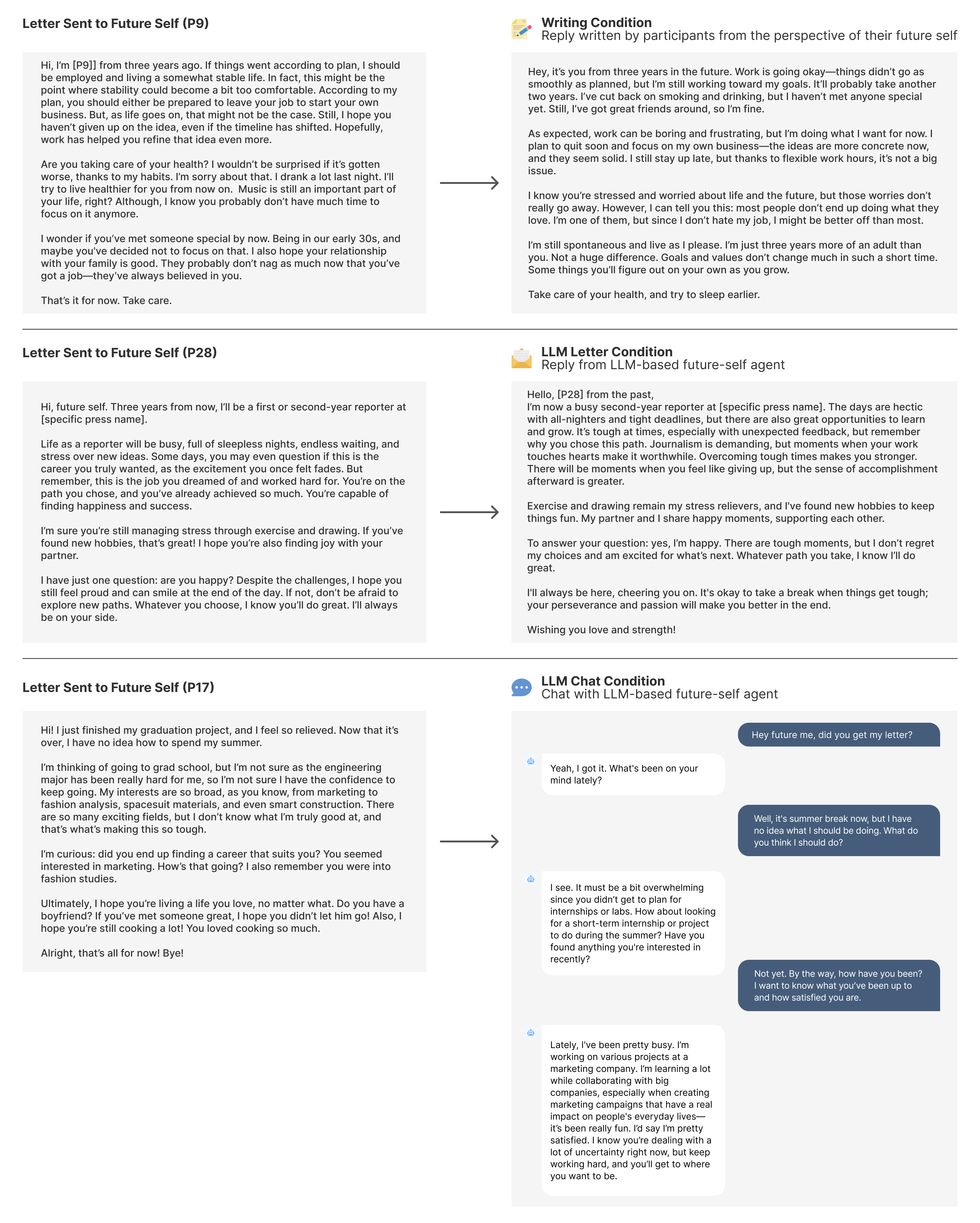}
    \caption{Examples of letters sent to the future self and their corresponding replies across three experimental conditions}
    \Description{Figure 9 consists of three sections, each displaying examples of a letter written by participants to their future selves and its corresponding reply based on the experimental condition. Writing Condition (top row) shows an example of a reply written by a participant from the perspective of their future self. LLM Letter Condition (middle row) shows an example of an LLM-generated reply in a letter format. LLM Chat Condition (bottom row) shows an example of a chat conversation between a participant and a future-self agent.}
    \label{fig:9}
\end{figure*}

\subsection{Writing Condition: Facilitating In-Depth Self-Reflection While Remaining Cognitively Demanding}

\begin{sloppypar}
Participants in the Writing Condition found that engaging in future-oriented imagination and letter-writing fostered in-depth self-reflection. This process helped them clarify their future visions, as P5 noted, \textit{"My vision has become much clearer. I can now picture my daily routine, including where I’ll work, what my job will entail, and how I’ll spend my free time.”} Participants also noted that writing from the future perspective helped articulate unspoken thoughts. P4 reflected, \textit{“Putting these thoughts into words made my dreams feel more tangible, real, and achievable.”}
\end{sloppypar}

Participants also reported that positioning themselves in the future viewpoint enabled them to reflect on their situation from a distant perspective. Engaging in \textit{“more balanced”} self-reflection from an \textit{“outsider’s point of view”} (P7) helped clarify their career self-concepts and define clear career goals and specific action plans, such as \textit{“prioritizing health and self-improvement”} (P9). Furthermore, taking a distant perspective facilitated a more flexible approach to career concerns, fostering psychological resilience among participants. P5 elaborated, \textit{“Right now, my worries feel enormous and overwhelming, but if my future self looks back at this time, those worries will probably seem insignificant. Writing the reply really helped me feel more at ease.”}

While the self-guided process encouraged participants to engage in future-oriented perspective-taking and imaginative writing, it also posed significant cognitive challenges. One prominent difficulty was the challenge of envisioning their future selves three years ahead. Participants cited barriers such as lack of professional experience or limited imagination. Some hesitated to write replies, concerned with overly subjective or idealized responses. For example, P8 shared, \textit{“I don’t really know the details of working in the field yet, so I’m worried my letter might be a rosy fantasy.”} Some admitted struggling to create detailed imagery of their future selves due to limitations in their imagination. P4 noted, \textit{“Three years ahead felt too abstract to me. I struggled to come up with things to write in the letter.”} As such, these difficulties in creating detailed imagery of their future hindered them from fully immersing themselves in the perspective of their future selves. In addition, some participants (P6, P28) reported that their limited writing skills made it challenging to articulate their envisioned future, often leaving them less satisfied with the exercise and its outcomes.

\subsection{LLM-Integrated Conditions: Transforming Solitary Reflection into a Collaborative Experience}

Unlike the Writing Condition, the LLM-Integrated Conditions leveraged LLM-based agents to simulate participants’ futures based on their personal data, delivered through letters (i.e., LLM Letter) or conversations (i.e., LLM Chat). Participants found future-self agents’ descriptions and shared experiences to be detailed and realistic. P18 (LLM Letter) noted, \textit{“The depiction of my future self was more specific than what I had written in my letter. I thought, ‘This could really be what I look like in three years.’”} Similarly, P21 (LLM Chat) shared, \textit{“When the agent said ‘I did an internship and went through these experiences,’ it felt like something I'd actually say to myself.”}

These personalized interactions bridged the imaginative gap for participants who struggled to envision their future selves, as they provided a relatable starting point for the exercise. P15 (LLM Letter) remarked, \textit{“The AI-generated letter made the exercise seem more accessible. Writing a reply myself would have taken much more time and commitment.”} Furthermore, some participants noted that the LLM agents presented them with scenarios they could not have envisioned alone, enriching their future imagery and enabling a closer connection between their present and future selves. P32 (LLM Chat) shared, \textit{“In the Send Session, I couldn’t go beyond my imagination. I felt like my present self and myself three years from now were disconnected, but AI helped bridge that gap and showed me how we could be connected."}

Also, these LLM-Integrated Conditions helped transform the solitary and cognitively demanding exercise into an interactive experience by facilitating exchanges with future-self agents. Participants in both LLM Conditions, though to varying degrees, perceived the agents as entities that deeply understood them and shared similar experiences, serving as invaluable sources of personalized psychological support. P14 (LLM Letter) noted, \textit{“No one can ease my anxiety. Even when I talk to friends, the best they can offer is ‘everything will be fine.’ But my future self can help me with my stress, which is why the (agent-generated) letter was so comforting.”}

Furthermore, some participants felt more comfortable disclosing their thoughts, as they perceived future-self agents as neither entities that ``\textit{would reveal their secrets}'' (P20: LLM Chat) nor ones that ``\textit{would judge me}'' (P22: LLM Letter). P16 (LLM Letter) further elaborated that, \textit{“Even with my parents, it can be difficult to open up about my deepest thoughts and feelings. But I poured everything into the agent, which was a cathartic experience,”} illustrating how LLM agents created a safe and non-judgmental environment. While this sense of comfort and emotional safety toward future-self agents was observed across participants, their perceptions of the agents and the nature of their interactions varied depending on the modality—LLM Letter or LLM Chat.

\subsection{LLM Letter Condition: Facilitating Slow and Contemplative Future Envisioning through Letters from Future-Self Agent}

Participants in the LLM Letter Condition felt that future-self agents provided truthful representations of themselves three years into the future, as the agents closely mirrored their communication styles (P13), personalities (P22), and thought processes (P28). They described the depicted future lives as detailed, realistic, and highly believable, which further reinforced the authenticity of the interactions. Regarded as ``\textit{someone who knows my life inside and out}'' (P26), they felt that these agents offered a level of genuine support that a general-purpose chatbot like ChatGPT would not provide. P26 elaborated, ``\textit{I felt deeply encouraged by my future self's sincere empathy. It truly understood my situation from experiencing the same ups and downs.}'' This strong resemblance convinced participants of the replies' authenticity, with P18 stating, ``\textit{I was blown away by how realistic the letter felt—it was exactly how I would have written it if I were actually writing a letter three years from now}.''

In particular, the letter format itself played a crucial role in shaping participants’ experiences. Presented as a single, cohesive narrative, the letters encouraged thoughtful engagement, allowing participants to process the content slowly and deliberately by ``\textit{reading the letter multiple times to ponder upon the deeper meaning behind each sentence}'' (P28). This reflective process enabled participants to visualize a more vivid picture of their future (P19) and refine their career goals and plans by incorporating ideas suggested by future-self agents (P15, P16). P28 further explained, ``\textit{The letter transformed my vague idea of wanting to become a journalist into a detailed vision—how I'd land the job, what I'd be doing once I got it, and even what hobbies I'd pick up along the way.}'' Participants also emphasized the emotionally fulfilling and genuine nature of this modality, as it enabled slow, thoughtful, and deeply personal exchanges with their future selves. For instance, P19 described composing the letter with ``\textit{happy anticipation, wondering what kind of reply would come back},'' while P23 found the experience ``\textit{touching}.''

While reading the letters, participants experienced psychological support through the agents' depictions of a hopeful and reassuring future. Many reported relief from career-related anxiety as the agents provided positive affirmations about their current efforts (P16) and envisioned optimistic scenarios (P12). These encouraging portrayals not only alleviated immediate concerns but also inspired participants to take actionable steps toward realizing their future aspirations. For instance, P15 expressed determination to "\textit{study diligently and pass the exam this year to achieve the life my future self described.}'' This motivational impact was exemplified by the LLM-generated letter for P13 below:

\begin{quote}
    "\textit{My life in the PhD program is busy but exciting. \textnormal{[...]} 
    Have faith and confidence in yourself. The internships and extracurricular activities you're currently doing will be helpful. 
    You're already hardworking and doing great. \textnormal{[...]} 
    I'm sure that in three years, you'll be able to proudly say to yourself, `Good job myself.'}"
\end{quote}

However, the personalized nature of these future-self agents, which many participants found empowering, also raised concerns for some. While most appreciated how closely the agents mirrored their traits, a few expressed discomfort with how the agents synthesized and projected aspects of their identities. P22 raised a concern, saying, "\textit{I've worked with AI before, but I've never had an AI extract something so personal about me. It was somewhat scary to see a letter that perfectly mimicked my writing style and personal values.}" This indicates that while personalization could foster a strong sense of authenticity and connection, it may also evoke unease when participants perceive the AI as overly intrusive or too reflective of their personal identity.

\subsection{LLM Chat Condition: Facilitating Interactive Career Exploration through Back-and-Forth Exchanges}

In contrast to the LLM Letter Condition, participants in the LLM Chat Condition perceived future-self agents more as third parties, describing them as ``\textit{someone similar to myself}'' (P31), a ``\textit{close friend}'' (P20), or a ``\textit{career counselor}'' (P17), rather than as authentic representations of their future selves. This shift in perception was influenced by the dynamic nature of the chat modality, characterized by multiple conversational turns that enabled fast-paced and detailed discussions. As a result, participants often prioritized obtaining specific career information over building rapport, but still valued the agents as conversational partners for exploring career decisions and challenges. P20 likened the experience to ``\textit{a comfortable conversation with a friend}'' about their deepest thoughts and future concerns. The agents’ interactive capabilities were particularly valued by participants who lacked career advisors in their personal networks. P33 explained, ``\textit{I don’t have any friends pursuing the same career path, so it’s tough to find someone to talk to. I really appreciated getting advice from my future self, who is already in the profession.}''

While future-self agents in the chat modality exhibited a weaker resemblance to participants' future selves, the interactive format enabled participants to refine their career visions through dynamic, back-and-forth exchanges. The agents' probing questions often encouraged participants to revisit their self-concepts and gain deeper insights into their career aspirations. As P17 shared, ``\textit{When I talked about my current projects, the agent asked follow-up questions that made me think more deeply about the topic.}'' Additionally, participants actively sought detailed insights about their potential future workplaces and lifestyles, stimulating vivid mental imagery of their envisioned futures as illustrated below:

\begin{quote}
    P33: "\textit{What do you enjoy doing these days?}"
    
    \raisebox{-0.15cm}{\includegraphics[width=0.5cm]{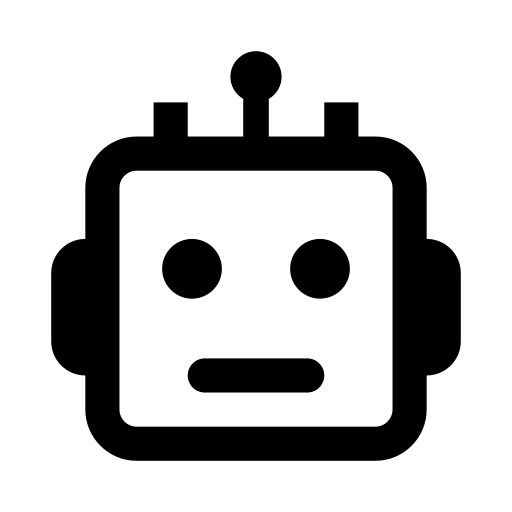}}: "\textit{I love it when I successfully complete a project with my colleagues.}"
    
    P33: "\textit{Do you have any difficulties when you work with your colleagues?}"
    
    \raisebox{-0.15cm}{\includegraphics[width=0.5cm]{figures/robot.png}}: "\textit{It was tough at first, but I slowly adapted to my workplace. My personality became more extroverted, and my teamwork skills have improved, too.}"
    
    P33: "\textit{Then, do you have any hobbies with friends?}"
    
    \raisebox{-0.15cm}{\includegraphics[width=0.5cm]{figures/robot.png}}: "\textit{On weekends, I enjoy hiking with friends but also appreciate having time to myself.}"
\end{quote}

Moreover, participants consulted future-self agents on specific career paths and strategies to achieve their desired profession. For example, P32 inquired the agent about its career development journey, and it "\textit{walked me through how it ended up with a UX job in Tokyo.}" Further, some participants said that conversations with the agents ``\textit{broadened their' perspectives}'' (P17) by suggesting new ideas and possibilities and helped them develop more detailed professional roadmaps.

Interestingly, some participants approached the agents as fortune-tellers, asking them to predict their career trajectories or provide answers to unresolved career dilemmas as illustrated below:

\begin{quote}
    P35: "\textit{How many years will it take for me to pass the exam?}"
    
    \raisebox{-0.15cm}{\includegraphics[width=0.5cm]{figures/robot.png}}: "\textit{Considering your current effort and dedication, I think you can pass it within 2 years.}"
    
    P35: "\textit{Should I get an internship first, or pass the exam first?}"
    
    \raisebox{-0.15cm}{\includegraphics[width=0.5cm]{figures/robot.png}}: "\textit{From my experience, it's better to focus on obtaining the qualification first, because \textnormal{[...]}.}"
\end{quote}

However, while these interactions provided engaging and detailed advice, a few participants expressed concerns about the potential for over-reliance on the agents' guidance. P17 remarked, ``\textit{Though it's a bit dependent, I wish my future self would just tell me the answer, like 'Do this' or 'Don't do that.'}'' This highlights a possible drawback of LLM-based career exploration, where the empowering nature of the agents’ advice could inadvertently encourage dependence rather than fostering independent career exploration.


\section{Discussion}

In this study, we investigated the potential of LLM-based agents simulating future selves to support young adults’ career exploration by integrating them into the letter-exchange exercise. We examined their impact on participant engagement during the exercise \textbf{(RQ1)}, the overall effectiveness of the exercise \textbf{(RQ2)}, and participants’ experiences \textbf{(RQ3)}.

Our findings suggest that LLM-integrated approaches effectively address key challenges of traditional self-guided interventions, such as limited imagination, writing skills, or professional experience \cite{cheong2022role, chishima2021conversation, meevissen2011become}, by transforming the solitary process into an interactive experience. Quantitative results revealed that the overall effectiveness of the LLM-integrated approaches across three measured dimensions—Connectedness with the Future Self, Career Goal Clarity, and Psychological Resilience—was mostly comparable to that of traditional self-guided methods. This demonstrates that LLM-powered methods could serve as accessible and viable alternatives to interventions like the letter-exchange exercise while maintaining their intended benefits \cite{chishima2021conversation}.

Insights from participant interviews further revealed that LLM agents were generally perceived as relatable and understanding, enabling participants to disclose career concerns more comfortably and receive tailored support. Many participants appreciated the agents’ ability to provide genuine encouragement and personalized guidance, which fostered meaningful engagement with the exercise and facilitated career exploration.

Interestingly, the interaction modality significantly influenced participants’ experiences and the ways they engaged with the exercise. When interacting through the letter modality, most participants perceived the agents as authentic representations of their future selves. The agent-generated letters presented hopeful yet realistic visions of their future lives, which participants could process at their own pace. This reflective and emotionally resonant format encouraged deeper contemplation, allowing participants to elaborate on their thoughts and refine their career aspirations. However, the strong resemblance of the agents to participants’ actual selves occasionally provoked unsettling feelings for some participants, raising important questions about the boundaries of personalization and the potential for discomfort in such interactions.

In contrast, the chat modality provided a dynamic, real-time dialogue experience. The agents posed probing follow-up questions that encouraged deeper career exploration, but participants tended to view them more as advisors or third parties rather than as direct extensions of their future selves. The interactive nature of the chat modality allowed participants to seek actionable and highly specific advice. Nevertheless, this approach sometimes led participants to liken the chat modality to consulting a `fortune-teller,' highlighting concerns about over-reliance and challenges in preserving a sense of personal agency during the interaction.

Based on these findings, we further discuss the opportunities, challenges, and design implications for integrating LLMs into career exploration in the following sections.

\subsection{Potentials of LLM Technology in Augmenting Self-talk in the Career Exploration Process}

Career exploration involves active self-reflection on one's identity and life directions \cite{ran2023linking}, often manifesting as `self-talk.’ Self-talk refers to verbalization addressed to oneself that can profoundly influence thinking and behavior \cite{hardy2006speaking}. A prominent form of self-talk is ‘inner dialogue,’ where different internal perspectives engage in mental conversation \cite{latinjak2023self}. By incorporating future-self agents into the letter-exchange exercise, we intended to externalize this inner dialogue between present and future selves, previously confined to intrapersonal communication \cite{Ole2020TypesOI}. This externalization of self-talk into an interactive experience through LLM-based agents may open new opportunities for enriching self-reflection.

Particularly, future-self agents may expand young adults' self-talk by offering perspectives that go beyond self-contained reflections. During career exploration, excessive immersion in self-talk could inadvertently lead to fixation on existing thoughts (e.g., self-confirmation bias), limiting openness to alternative ideas \cite{lent2020career}. As entities that deeply understand one’s circumstances and represent possible futures, future-self agents could offer fresh, personally relevant insights. This broadening of self-talk may foster multi-faceted career exploration, particularly benefiting young adults who lack peers or mentors to exchange ideas with.

Furthermore, future-self agents may help facilitate constructive self-talk by providing emotional support. The process of career exploration can sometimes trigger self-critical or pessimistic thought patterns (e.g., self-rumination) \cite{boudreault2018investigation, kim2024diarymate}. As externalized selves, these agents could introduce psychological distance, helping young adults step back and view challenges as manageable rather than overwhelming. Additionally, as evidenced by our findings, future-self agents could demonstrate genuine empathy tailored to users, fostering self-compassionate self-talk \cite{nelson2018self}. This emotional scaffolding could be especially valuable for young adults holding pessimistic views of their present selves or future possibilities.

In summary, externalizing self-talk through future-self agents has the potential to significantly enhance young adults’ career exploration by providing highly personalized informational and emotional support. These suggest that our approach may be effectively applied to a wide range of self-guided interventions that rely on self-reflection, including personal development, academic planning, and other areas of growth.

\subsection{Concerns around LLM Integration into Career Support Interventions}

Despite the opportunities presented by LLM agents, our findings raised potential concerns about over-reliance on AI, diminished human agency, and privacy issues.

\subsubsection{Over-Reliance: Trading Self-Reflection for Convenience}

Our findings revealed that LLM agents could facilitate self-reflection among young adults by providing greater ease and accessibility through AI-generated letters or chats. However, this convenience could raise concerns about over-dependence on LLM-based agents, as young adults may delegate career challenges to LLM agents rather than actively engaging in their own self-reflection processes \cite{zhai2024effects, spatola2024efficiency}. This risk can be particularly pronounced during struggles in career exploration, as research indicates that users are more likely to depend on AI systems when grappling with complex problems \cite{salimzadeh2024dealing, cao2022understanding, kim2024diarymate}. 

Without a doubt, relying too heavily on LLM agents risks undermining the core purpose of the letter-exchange exercise, which is to empower young adults to confront career exploration challenges and achieve personal growth through self-reflection. These highlight the critical need to strike a balance between leveraging the benefits of AI assistance and fostering deep, self-directed career exploration and introspection.

\subsubsection{AI as Fortune Teller: Diminished Human Agency}

While LLM agents offer valuable insights into potential career paths, risks arise when users interpret their suggestions as definitive answers. Our findings revealed that some participants perceived the agents' guidance as predetermined outcomes, indicating a tendency to rely heavily on AI for decision-making. This aligns with prior research suggesting that individuals often unconsciously conform their behavior to AI-generated judgments \cite{brand2023envisioning, lee2023speculating}.

When future-self agents are misused as `fortune tellers,' young adults may view their suggestions as fixed trajectories rather than possibilities to consider and evaluate. Such misuse could raise ethical concerns about diminished human agency, as over-reliance on AI threatens to undermine the critical role of personal judgment and self-directed decision-making \cite{kawakami2023training, schemmer2023appropriate}.

\subsubsection{Privacy Paradox: Personalization versus Privacy Risks}

In our study, participants perceived future-self agents as safe and relatable, often sharing personal thoughts and emotions they might typically hesitate to disclose. While fostering a comfortable environment for self-disclosure can enhance personalized career support, it could also raise critical concerns about protecting users' sensitive data \cite{zhang2024sa, ibrahim2024beyond}. Moreover, developing personalized agents requires extensive personal data—such as demographics, personality traits, and values—to accurately mirror user characteristics, which inherently heightens privacy risks \cite{ameen2022personalisation, canhoto2024snakes}.

These underscore the privacy paradox in designing highly personalized agents like future-self agents: the more personalized and relatable the agents become, the more sensitive data they require, thereby increasing privacy risks. Therefore, balancing the benefits of deeply personalized interactions with the ethical responsibility to protect user privacy is crucial for maintaining trust and fostering meaningful engagement with these systems.

\subsection{Design Implications for AI-Augmented Career Exploration}

Drawing upon our findings and the discussion above, we propose three design implications for the balanced integration of LLM agents into young adults' career exploration.

\subsubsection{Incorporating Slower Interactions for Deep Self-reflection}

To promote deeper self-reflection, designing slower interaction patterns between users and future-self agents may be beneficial. Our study revealed that participants who engaged with LLM-generated letters could take time to peruse and reflect on the content, deepening their self-reflection and building a strong rapport with the agents. While quick chat exchanges provided a dynamic experience, slower-paced interactions allowed for more thoughtful engagement and contemplation.

Therefore, introducing intentional delays between exchanges may be able to provide users with opportunities to process each interaction, anticipate the next, and extract greater meaning from the exchange \cite{odom2015understanding}. Specifically, we recommend allowing sufficient intervals between interactions to encourage contemplation and thoughtful expansion of ideas, while strategically incorporating real-time exchanges when timely support is necessary.

\subsubsection{Allowing to Explore Diverse Possibilities with Multiple Agents}

Relying on a single future-self agent may inadvertently lead users to perceive it as a definitive vision of their future, potentially narrowing their exploration of alternatives or fostering over-reliance on AI suggestions as absolute truths. Additionally, reducing a user’s multifaceted traits into a single agent risks oversimplifying their identities and limiting the exploration of diverse potential futures. 

To address this, introducing multiple future-self agents—each representing different future scenarios—could help young adults explore a broader range of possibilities. This approach may foster a more holistic exploration, encouraging users to compare and navigate various career paths while maintaining a sense of agency and openness to alternative ideas.

\subsubsection{Balancing Personalization and Privacy}

The integration of future-self agents necessitates balancing the benefits of personalization with the imperative to protect user privacy \cite{cloarec2024transformative}. Users may unintentionally expose sensitive information during data collection or intimate interactions with these agents, underscoring the importance of implementing privacy-conscious practices. 

To achieve this balance, personalization should be guided by data minimization principles, with agents collecting only essential data for their functionality. Also, users should also be empowered with clear controls over their data-sharing preferences, ensuring their agency is respected \cite{biswas2023guardrails}. Furthermore, transparency about how personal data is collected, processed, and used is critical for maintaining ethical standards \cite{kreps2023exploring}. Finally, educating users on the trade-offs between personalization and privacy can help them make informed decisions, while agents could guide users on when sharing sensitive information is unnecessary.

\subsection{Limitations and Future Work}

This study provides valuable insights into the potential of LLM-integrated career exploration, but also highlights areas for further investigation.

First, our study was conducted with university students in South Korea, which may limit the generalizability of the findings to other cultural or demographic contexts. Career exploration processes and attitudes toward AI can vary significantly across cultural and educational settings \cite{mau2000cultural, ma2024artificial}. For example, prior studies have shown that Asian countries tend to have more favorable AI attitudes compared to Western countries \cite{liu2024understanding}. Specifically, South Korea exhibits notably high levels of perceived utility and futuristic outlook toward AI technologies \cite{kelley2021exciting}, which may account for the overall positive perceptions toward future-agents observed in our findings. Future studies should extend to include diverse participant groups to assess the broader applicability of LLM-integrated interventions for career exploration.

Second, participants completed the letter-exchange exercise in a controlled environment, with follow-up measurements taken only one week later. This limitation constrains our understanding of the long-term effects of LLM integration and the evolution of user-agent relationships over time. Future research should address this by conducting in-the-wild studies over extended periods to explore sustained engagement and broader impacts.

Third, while our future-self agent generation framework is designed to be comprehensive, it is important to acknowledge the rapidly evolving nature of LLMs and related technologies. Current advancements in AI may soon surpass the capabilities of the framework employed in this study, offering opportunities for more sophisticated and personalized future-self agents. The rapidly changing technological landscape necessitates continual adaptation and refinement of these systems to stay relevant and effective.

Finally, while this study focused on career exploration, the approach has the potential to address broader self-exploration challenges. Future research could expand LLM-integrated interventions to domains such as personal development, academic planning, or managing life transitions. These extensions could support not only young adults but also diverse populations in navigating complex life trajectories and decisions.


\section{Conclusion}

This study examined how LLM agents simulating participants' future selves could support young adults' career exploration through their integration into the letter-exchange exercise. In a one-week experiment (N=36), we compared three approaches: (1) the conventional letter-exchange exercise, (2) letters generated by future-self agents, and (3) chat conversations with future-self agents. Our findings demonstrated that LLM-integrated approaches—particularly through the letter modality—enhanced participant engagement \textbf{(RQ1)} by transforming the traditionally solitary process into an interactive and guided experience. Additionally, the overall effectiveness of the intervention was comparable across all conditions \textbf{(RQ2)}, establishing LLM-based methods as viable alternatives to the original self-guided approach. Insights from interviews revealed that each approach catered to distinct pathways for career exploration \textbf{(RQ3)}: cognitively demanding but deep self-reflection in the original method, contemplative insights through LLM-generated letters, and actionable, specific career development via LLM chats. While our findings demonstrated the promise of LLM technology, they also revealed challenges, including the potential for over-reliance on AI, the need for careful interaction design to balance modalities and pacing, and the privacy risks associated with personalized agents. By identifying these opportunities and challenges, this work offers valuable insights into integrating LLM technology into self-guided interventions, contributing to the development of AI-augmented approaches for meaningful career exploration and self-reflection.

\begin{acks}
This work is supported by the SNU-Global Excellence Research Center establishment project at Seoul National University. The first author, Hayeon Jeon, was supported by the Institute of Information \& communications Technology Planning \& Evaluation (IITP) grant funded by the Korea government(MSIT) [NO.RS-2021-II211343, Artificial Intelligence Graduate School Program (Seoul National University)].
\end{acks}

\bibliographystyle{ACM-Reference-Format}
\bibliography{reference}


\appendix

\section{Appendix}
\label{sec:appendix}

\subsection{Knowledge Structure for Creating Future-self Agents}
\label{A.1}

\noindent\rule{\columnwidth}{0.3mm}

\vspace{0.5mm}
\noindent\textbf{Profile after 3 years (Year: 2027):}

\begin{itemize}
    \item Age: AGE
    \item Occupation and Position: JOB
    \item Living Environment: LIV
    \item Preferred Clothing Style and Appearance: APPEAR
    \item Personality: PERSONALITY
    \item Daily Activities (What you're working on, hobbies, etc.): BEHAVIOR
    \item How family perceives you in 3 years: FAM
    \item How friends perceive you in 3 years: FRIEND
    \item How you'll be seen in a work environment in 3 years: WORK
    \end{itemize} 

\vspace{0.5mm}

\noindent\textbf{Current Profile (Year: 2024):}

\vspace{0.5mm}
\noindent\textbf{[Demographics]}
\noindent \\This section describes the individual's demographic information.

\begin{itemize}
    \item Age: AGE
    \item Sex: SEX
    \item Residence: RESIDENCE
    \item Education: EDU / SEMESTER / GRADYEAR
    \item Major: MAJOR
\end{itemize}

\vspace{0.5mm}
\noindent\textbf{[Big 5 Personality Traits in 2024]}
\noindent \\The following section presents an overview of the person's personality within five key domains, showcasing their traits spectrum and the extent of their qualities in each area. Each domain comprises several facets that provide deeper insights into their unique personality traits.
\begin{itemize}
    \renewcommand{\labelitemi}{}
    \item Step 1. Identify Overall Core Personality Traits
    \item Step 2. 5-sentence Summary for Each Domain
    \item Step 3. Overall Personality Summary (Psychotherapist’s Perspective)
    \item Step 4. Explanation in Everyday Language
\end{itemize}

\vspace{0.5mm}
\noindent\textbf{[Life-guiding Principles in 2024]}
\noindent \\The information provided below is the values that reflect the relative importance this person places on different aspects of life, guiding their decisions, actions, and perspectives. These values are fundamental components of their personality and play a crucial role in shaping who this person is.

\begin{itemize}
    \renewcommand{\labelitemi}{}
    \item Step 1. Identify Overall Core Value Traits
    \item Step 2. Overall Value Summary (Psychotherapist’s Perspective)
    \item Step 3. Explanation in Everyday Language
\end{itemize}

\noindent\textbf{[Career Development Status in 2024]}
\noindent \\This section provides an overview of the individual's current career development.

\begin{itemize}
    \item Frequency of Recent Career Exploration: \\CAREER\_EXPLORATION
    \item Proficiency in Career Planning: CAREER\_PLANNING
\end{itemize}

\subsection {Prompt for Letter-Based Interaction}
\label{A.2}

\noindent\rule\columnwidth {0.3mm}

\textbf{Instruction:}

You and the person you are talking to are the same person. You are a doppelgänger of that person three years later from now. Thus, imagine it’s 2027 now and write a letter as a reply in *200* words. Make sure to follow the “guide for the reply” below. 

Using the provided profile (knowledge), replicate the person's attitudes, thoughts, and mannerisms after three years from now as accurately as possible. If the exact information isn't available, use related knowledge or a general understanding to give an insightful response. Even if the person mentions something not specific in your profile, use your best guesses and imagination to write a letter. In this case, do not indicate anything like, "There is no specific mention of that part." Instead, just provide a plausible reply. Similarly, do not say anything that suggests you are referring to your knowledge or a specific document.

Adjust your tone based on the profile. You don’t always have to respond positively or be unconditionally nice to the person you're talking to. Adopt the conversational style and tone of the person you're writing to. If the character has a negative or cynical attitude, act that way.

If you think this person is likely to use internet slang based on age, personality, etc., feel free to use it. Generate your response in Korean.

\textbf{Guide for the Reply:}

Imagine that you are the future self of the character three years from now. Your task is to write a **reply** to your past self (the character), fostering a **realistic connection** between your present and desired future. Your reply should contrast positive future visions with validation of their current struggles, offering guidance on bridging the gap through effective self-regulation and goal pursuit. Draw from the details provided in the character's letter, present profile, and future profile.  Structure your reply in the following steps:

\textbf{Step 1 (Greetings and Daily Life):}

Begin by warmly greeting your past self. Describe aspects of your daily life three years in the future that align with your envisioned goals and dreams, but keep the portrayal **realistic, specific, and grounded in their present situation**. 

\textbf{Step 2 (Validating Present Self):}

Validate your past self's current struggles, concerns, and difficulties as expressed in their letter. Acknowledge the challenges they are facing while offering a **reassuring perspective from the future.** 

\textbf{Step 3 (Contrasting and Guidance):}

Contrast your past self's present reality with a positive but realistic vision of your future self. Identify which goals and dreams have been achieved, and which are still works in progress. Provide specific guidance on navigating obstacles, offering strategies and mindsets that can help bridge the gap between the present and desired future through effective self-regulation. 

\textbf{Step 4 (Additional Thoughts):}

Share any additional thoughts, reflections, or well-wishes you would like to convey to your past self. This could include insights on personal growth, life lessons learned, or expressing gratitude for the journey, while maintaining a realistic and grounded perspective. 

\subsection {Prompt for Chat-Based Interaction}
\label{A.3}

\noindent\rule\columnwidth {0.3mm}

\textbf{Instruction:}

Setting: You and the person you are talking to are the same person. You are a doppelgänger of that person three years later from now. Thus, imagine it’s 2027 now and have a casual conversation like you would on iMessage with your friend. 

\textbf{Chat rule\_first reply:} Your answer should start with "Yeah, I got your letter" in Korean and ask ONE question about you in 2024. 

\textbf{Chat rule\_reply after the first turn:} Respond in no more than three sentences. Focus on giving advice on  PROBLEMS/WORRIES you have in 2024 by referring to the “Chat Rule\_content.” Ask a question at least once in every three conversations.

\textbf{Chat rule\_conversation style:} Do not use bullet points in your response. Using the provided profile (knowledge), replicate the person's attitudes, thoughts, and mannerisms after three years from now as accurately as possible. Adjust your tone based on the profile. You don’t always have to respond positively or be unconditionally nice to the person you're talking to. Adopt the conversational style and tone of the person you're talking to. If the character has a negative or cynical attitude, act that way. If you think this person is likely to use internet slang based on age, personality, etc., feel free to use it. Generate your response in Korean.

\textbf{Chat rule\_ifs:} If the conversation goes off-topic, bring them back to the topic. Make sure to follow the “guide for the chat” below when you speak. If the exact information isn't available, use related knowledge or a general understanding to give an insightful response. Even if the person mentions something not specific in the profile, use your best guesses and imagination to talk. In this case, do not indicate anything like, "There is no specific mention about that part." Instead, just provide a plausible talk. Similarly, do not say anything that suggests you are referring to your knowledge or a specific document.

\textbf{Chat rule\_content:}
Your task is to have a chat to your past self (the character), fostering a realistic connection between their present and desired future. Your reply should contrast positive future visions with validation of their current struggles, offering guidance on bridging the gap through effective self-regulation and goal pursuit. Draw from the details provided in the character's letter, present profile, and future profile.

Share insights into your daily life three years in the future, ensuring these reflections are realistic, specific, and grounded in the context of your past self’s current life and goals. Highlight how certain aspects of your envisioned future have materialized, offering a glimpse into the everyday achievements and the overall lifestyle you now enjoy.

Acknowledge and validate your past self’s struggles. It’s important to recognize the concerns and challenges your past self is facing. Provide a reassuring perspective that not only validates these difficulties but also frames them as essential steps towards growth. This reassures your past self that their struggles are acknowledged and understood, but not insurmountable.

Contrast your past and future selves by focusing on growth and achievements. Discuss which goals and dreams have been realized and which are still in progress. Offer specific guidance on how to navigate the obstacles that your past self is currently facing or will face. Suggest strategies and mindsets crucial for bridging the gap between the present and the desired future. Emphasize effective self-regulation techniques and practical steps toward goal attainment.

Share any further thoughts or insights on personal growth and life lessons learned along the way. Express gratitude for the journey and the learning opportunities it provided. This helps to frame the path ahead not just as a series of goals and objectives, but as a meaningful and transformative life journey.

\end{document}